\documentclass[fleqn,usenatbib,letterpaper]{mn2e}
\usepackage{graphicx}
\usepackage{amsmath}
\usepackage{array}
\newcolumntype{L}[1]{>{\raggedright\let\newline\\\arraybackslash\hspace{0pt}}m{#1}}
\def\jnl@style{\it}
\def\aaref@jnl#1{{\jnl@style#1}}

\def\aaref@jnl#1{{\jnl@style#1}}

\def\aj{\aaref@jnl{AJ}}                   
\def\araa{\aaref@jnl{ARA\&A}}             
\def\apj{\aaref@jnl{ApJ}}                 
\def\apjl{\aaref@jnl{ApJ}}                
\def\apjs{\aaref@jnl{ApJS}}               
\def\ao{\aaref@jnl{Appl.~Opt.}}           
\def\apss{\aaref@jnl{Ap\&SS}}             
\def\aap{\aaref@jnl{A\&A}}                
\def\aapr{\aaref@jnl{A\&A~Rev.}}          
\def\aaps{\aaref@jnl{A\&AS}}              
\def\azh{\aaref@jnl{AZh}}                 
\def\baas{\aaref@jnl{BAAS}}               
\def\jrasc{\aaref@jnl{JRASC}}             
\def\memras{\aaref@jnl{MmRAS}}            
\def\mnras{\aaref@jnl{MNRAS}}             
\def\pra{\aaref@jnl{Phys.~Rev.~A}}        
\def\prb{\aaref@jnl{Phys.~Rev.~B}}        
\def\prc{\aaref@jnl{Phys.~Rev.~C}}        
\def\prd{\aaref@jnl{Phys.~Rev.~D}}        
\def\pre{\aaref@jnl{Phys.~Rev.~E}}        
\def\prl{\aaref@jnl{Phys.~Rev.~Lett.}}    
\def\pasp{\aaref@jnl{PASP}}               
\def\pasj{\aaref@jnl{PASJ}}               
\def\qjras{\aaref@jnl{QJRAS}}             
\def\skytel{\aaref@jnl{S\&T}}             
\def\solphys{\aaref@jnl{Sol.~Phys.}}      
\def\sovast{\aaref@jnl{Soviet~Ast.}}      
\def\ssr{\aaref@jnl{Space~Sci.~Rev.}}     
\def\zap{\aaref@jnl{ZAp}}                 
\def\nat{\aaref@jnl{Nature}}              
\def\iaucirc{\aaref@jnl{IAU~Circ.}}       
\def\aplett{\aaref@jnl{Astrophys.~Lett.}} 
\def\apspr{\aaref@jnl{Astrophys.~Space~Phys.~Res.}}
\def\bain{\aaref@jnl{Bull.~Astron.~Inst.~Netherlands}} 
\def\fcp{\aaref@jnl{Fund.~Cosmic~Phys.}}  
\def\gca{\aaref@jnl{Geochim.~Cosmochim.~Acta}}   
\def\grl{\aaref@jnl{Geophys.~Res.~Lett.}} 
\def\jcp{\aaref@jnl{J.~Chem.~Phys.}}      
\def\jgr{\aaref@jnl{J.~Geophys.~Res.}}    
\def\jqsrt{\aaref@jnl{J.~Quant.~Spec.~Radiat.~Transf.}}
\def\memsai{\aaref@jnl{Mem.~Soc.~Astron.~Italiana}}
\def\nphysa{\aaref@jnl{Nucl.~Phys.~A}}   
\def\physrep{\aaref@jnl{Phys.~Rep.}}   
\def\physscr{\aaref@jnl{Phys.~Scr}}   
\def\planss{\aaref@jnl{Planet.~Space~Sci.}}   
\def\procspie{\aaref@jnl{Proc.~SPIE}}   

\begin{document}
\title[On the dark matter profile in Sculptor]{On the dark natter profile in Sculptor: breaking the $\beta$ degeneracy with Virial shape parameters}%
\author[T. Richardson \& M. Fairbairn]{Thomas Richardson\thanks{thomas.d.richardson@kcl.ac.uk}$^{1}$, Malcolm Fairbairn\thanks{malcolm.fairbairn@kcl.ac.uk}$^{1}$\\$^{1}$Physics, Kings College London, Strand, London WC2R 2LS, UK}
\maketitle
\begin{abstract}
We present a new method for studying tracers in gravitational systems where higher moments of the line-of-sight velocity distribution are introduced via Virial equations rather than the Jeans equations. Unlike the fourth order Jeans equations, the fourth order Virial equations can simply be added to the standard second order Jeans equation without introducing a new anisotropy parameter $\beta^{\prime}$. We introduce two new \emph{global} shape parameters $\zeta_A$ and $\zeta_B$ which replace the kurtosis as a more statistically robust measure of the shape of the line of sight velocity distribution.  We show that in the case of stars in dwarf spheroidal galaxies these new parameters can significantly reduce the range of density profiles that are otherwise consistent with the observed stellar kinematics (a problem sometimes known as the $\beta$ degeneracy). Specifically, we find that $\zeta_A$ focuses tightly on a subset of solutions where cusped density profiles are degenerate with more concentrated cored dark matter haloes. If the number density of stars $\nu(r)$ is fixed, then introducing $\zeta_B$ can further reduce the space of solutions by constraining the outer slope of the dark matter density profile. Assuming a Plummer profile for $\nu(r)$ we recover the surprising result that the dark matter in Sculptor may be cuspy after all, in contrast to the conclusions of other approaches.
\end{abstract}


\section{Introduction}

The $\Lambda$ cold dark matter ($\Lambda$CDM) paradigm suggests that structure formation should be hierarchical and that each galactic dark matter halo such as the Milky Way should have around it a family of increasingly smaller sub haloes, the larger of which retain some baryons and become dwarf spheroidal galaxies.  These smaller haloes have relatively shallow gravitational potentials and therefore retain few baryons, making their mass to light ratio very large compared to more massive haloes. This makes them excellent laboratories for studying dark matter.

Understanding the density profile of dwarf spheroidal galaxies can put constraints on the self annihilation of dark matter \citep{bergstrom} or the possibility of dark matter self-interaction \citep{spergel}.  It can also give us information about the way in which baryonic feedback can change the shape of dark matter potentials \citep{pontzen}.  There have been claims over the years that there are fewer dwarf spheroidals observed orbiting the Milky Way than we expect in the $\Lambda$CDM model \citep{toofew,toobigtofail}.  The reason for this maybe due to a lack of understanding of baryonic effects upon dark matter haloes or something more fundamental.  Recently it has been suggested again that alternatives to $\Lambda$CDM may be a better fit to the data, in particular because dwarf spheroidal density profiles do not seem to be cuspy as predicted in cold dark matter models but more cored \citep{walkerloeb}. These results add to evidence for cored dark matter haloes in other systems such as low surface brightness galaxies \citep{LSBcores} and dwarf spiral galaxies \citep{spiralcores}. Indeed, \cite{gentile} claimed that neither spherical nor triaxial cusped haloes can describe the observed kinematics in dwarf spiral DDO 47. It is therefore critical to try and understand the shape and depth of the gravitational potential in dwarf spheroidal galaxies as the results have a strong bearing on many fundamental questions.

The traditional way of obtaining the gravitational potential is to fit the variance of the line of sight (LOS) velocity dispersion of stars to candidate potentials using the second order moments of the Collisionless Boltzmann Equation, also known as the Jeans equation.  Unfortunately, this method has degeneracies as different choices for the stellar velocity anisotropy parameter,
\begin{equation}
\label{anis}
\beta(r) \equiv 1-\frac{\sigma^{2}_{t}(r)}{2\sigma^{2}_{r}(r)},
\end{equation}
where $\sigma^2_r$ and $\sigma^2_t=2\sigma^2_\theta=2\sigma^2_\phi$ represent the variance of the radial and tangential velocity distributions respectively, imply different gravitational potentials. Since we cannot measure $\beta$, we cannot ultimately know what the gravitational potential is \citep{Wolf}.

Because of the recent improvements in data sets for dwarf spheroidals there has recently been a number of new dynamical methods that supersede the standard Jeans analysis. Metallicity information for stars in dwarf spheroidals can separate them into distinct stellar sub-components \citep{battagliamulti,walkdat,amoriscomulti}. By using the Jeans equation to simultaneously fit two stellar sub-populations to the same potential, \cite{battaglia} were able to break the $\beta$ degeneracy to some extent in Sculptor and found that rising anisotropy models and cored density profiles were favoured. Subsequent studies with the projected Virial theorem \citep{amorfornax,evansvirial} and $\beta$-independent mass estimators \citep{penarrubia} were able to exclude cuspy density profiles in Fornax and Sculptor with a high statistical significance. Use of the projected Virial theorem on multiple tracer populations of globular clusters has also recently been applied  to nearby elliptical galaxies \citep{agelip}. The simplicity of these analytic methods coupled with claims \citep{chervin} that they can be successful on simulations of triaxial haloes (though see \cite{kowal} for a challenge of this result) gives compelling evidence for cored profiles in dwarf spheroidals.  

In spite of this, a lack of independent evidence from alternative single-component approaches to support these claims has left the cusp/core debate in dwarf spheroidal galaxies open. \cite{strigkurt} claimed that DM haloes from the Aquarius simulation can provide an acceptable single-component fit to the photometric and kinematic data in Milky Way dwarf spheroidals. Other single-component analyses with Bayesian hierarchical modelling \citep{martinez} and Schwarzschild orbit-based modelling \citep{Breddelsall,jardelnonpar} find no strong evidence for universal cusps or cores when considering the Milky Way's dwarf spheroidals as a whole. In contrast to these sophisticated techniques, our approach is to try and find the simplest analytic method possible that can break the $\beta$ degeneracy with a single tracer component and without a loss of generality.   
      
By fitting to the kurtosis of LOS velocity data in addition to the velocity dispersion, there is an extra constraint that can distinguish between different fits to the standard Jeans equation. Joint fits to the dispersion and kurtosis with second and fourth order Jeans equations \citep{merry} have been conducted on galaxy clusters \citep{mamoncoma}, dwarf spheroidals \citep{Lokas05} and recently giant elliptical galaxies \citep{ellip}. State-of-the-art observations of Milky Way dwarf spheroidal galaxies \citep{batdat,walkdat} now offer LOS velocities and projected radii for thousands of tracer stars. As data sets continue to expand, the inclusion of higher velocity moments is not only possible but also useful, especially if deviations from Gaussianity are statistically significant [as they are in Sculptor \citep{Amorisco}].  Previous work \citep{me} demonstrated that in general analogues to the Jeans equations for fourth moments of the velocity distribution \citep{merry} cannot be solved with the anisotropy parameter $\beta$ and gravitational potential alone. For a unique solution we must specify a new parameter such as
\begin{equation}
\beta^{\prime} = 1-\frac{3}{2}\frac{\langle v^2_rv^2_t\rangle}{\langle v^4_r \rangle}
\end{equation}
that relates fourth order moments $\langle v^4_r \rangle$ and $\langle v^2_rv^2_t\rangle = 2\langle v^2_rv^2_\theta\rangle= 2\langle v^2_rv^2_\phi\rangle$ and therefore allows us to characterize the relative shape (i.e. relative kurtosis as opposed to relative dispersion) of the radial and tangential velocity distributions.  Efforts to break the classical degeneracy with $\beta$ are therefore compromised by a new degeneracy of solutions for different choices of $\beta^{\prime}$.  

In this work we consider a simpler solution that is not compromised in this way. Higher order analogues of the projected Virial theorem $2 K_z + W_z = 0$ have been derived \citep{merry,kent} that give global constraints on velocity moments by effectively integrating the spherical Jeans equations over all radii.  While this results in a loss of spatial information it can be used instead of the Jeans equations to find out what density profiles are compatible with the LOS velocity data. Interestingly, fourth moment analogues to the projected Virial theorem do not depend on the relative {\it shape} of the radial and tangential velocity distributions $\beta^{\prime}$ (a fourth moment quantity) but only the relative {\it widths} $\beta$ (a second moment quantity) and the gravitational potential $\Phi$. As such the potential terms for fourth order projected Virial theorem estimators can be evaluated with the same parameter set as the second order Jeans equation without introducing new parameters such as $\beta^{\prime}$.  In essence, the fourth order Virial equations give us two extra constraints on the LOS data for free and without loss of generality. It has been shown that this result extends to $n\rm{th}$ order \citep{kent} such that $n$ Virial equations for moments of the LOS velocity distribution $\langle v^{2n}_{z}\rangle$ are independent of the $n$th order anisotropy parameter $\beta_n$ (see \cite{me} for definition) and depend only on parameters $\beta_j$ with $j<n$. 

In this work, we will investigate the use of higher order analogues of the projected Virial theorem in conjunction with conventional Jeans analyses to try and break the classic degeneracy problem.  In \S2, we will introduce two new shape parameters $\zeta_A$ and $\zeta_B$ which are constructed from the fourth order projected Virial theorems and which replace the kurtosis as a measure of the shape of the LOS velocity distribution that can be estimated from the data. We test the statistical properties of these Virial shape parameters with simulated data and show that these new measures seem to be more robust to systematics than the kurtosis in radial bins.

In \S3, we will perform an illustrative analysis on Sculptor with constant $\beta$ to highlight the key properties of the Virial shape parameters and to gain an intuition for how they can break the $\beta$ degeneracy. Finally, we do a generalized analysis in \S4 with non-constant $\beta$ to see which density profiles are consistent with the Jeans equation and both fourth order projected Virial theorems. Implications of key assumptions such as spherical symmetry are discussed in \S5.
 
\section{Virial Shape Parameters}

In this section, we provide formulae for the dynamics in our analysis. We state fourth order projected Virial theorems first derived in \cite{merry} that will replace the fourth order Jeans equations as conditions on the fourth moments of the LOS velocity distribution from the Collisionless Boltzmann equation. Whereas the fourth order Jeans equations describe the fourth moments locally as a function of radius, the Virial equations describe global averages. Using the Virial equations we define new shape parameters $\zeta$ that are global analogues to the kurtosis of the LOS velocity distribution. To replace the sample kurtosis in radial bins we define simple estimators of the $\zeta$ parameters (which we denote $\widehat{\zeta}$) that can be calculated trivially from discrete LOS velocity data. The statistical properties of these estimators are discussed at the end of the section.       

\subsection{Projected Virial theorem at second and fourth order}
Under the assumptions of spherical symmetry \citep{merry} and a constant mass-luminosity ratio for the stars $\Upsilon_\star$, the projected Virial theorem $2 K_z + W_z = 0$ yields the following relation,
\begin{equation}\label{PVT}
\int^{\infty}_{0} \Sigma \langle v^2_z \rangle R \rm{d}R = \frac{2}{3} \int^{\infty}_{0} \nu \frac{ \rm{d}\Phi }{\rm{d}r} r^3 \rm{d}r.
\end{equation}
where $r$ is the (3D) physical radius of each star from the centre of the galaxy, $\nu(r)$ is the (3D) number density of stars, $R$ and $\Sigma(R)$ are projections of $r$ and $\nu$ on to the plane perpendicular to the LOS and $\Phi$ is the total gravitational potential. With our assumption of a constant mass-luminosity ratio $\Upsilon_\star$ for the stars then $\rm{d}\Phi/\rm{d}r = GM(r)/r^2$ and the total (stars+dark matter) enclosed mass of the system at radius $r$ is given by,
\begin{equation}
M(r) = 4\pi \int^{r}_{0} s^2 (\Upsilon_\star \nu(s) + \rho(s)) ds,
\end{equation}
where $\rho(r)$ is the mass density of dark matter.

We denote the $2n$th moment of the radial velocity distribution (functions of $r$) by $\langle v^{2n}_r \rangle$. For a spherically symmetric system the phase space distribution function is simply the 3D velocity distribution as a function of radius $f(r,v_r,v_\theta,v_\phi)$ and the radial velocity moments are thus defined as,
\begin{equation}
\nu(r)\langle v^{2n}_r \rangle(r) = \int v^{2n}_r f(r,v_r,v_\theta,v_\phi) d^3\textbf{v}.
\end{equation}
We will often choose to drop the explicit $r$ dependence of the radial velocity moments for brevity. Due to the distance of dwarf spheroidal galaxies we currently only have access to spectroscopic measurements of each stars velocity along the LOS and to the projected position $R$ of each star on the sky. Instead of the full phase space distribution we therefore consider an observable projected phase space distribution $f_z(R,v_z)$ that describes the LOS velocity distribution as a function of the projected radius $R$. Here, the $z$ indicates that the $z$ axis is chosen as the LOS direction so that $\langle v^{2n}_{z}\rangle$ are moments of the LOS velocity distribution (as a function of $R$),
\begin{equation}
\Sigma(R)\langle v^{2n}_z \rangle(R) = \int v^{2n}_z f_z(R,v_z) dv_z.
\end{equation}
The number density of stars as a function of projected radius $\Sigma(R)$ can be calculated from the 3D number density $\nu(r)$ via,
\begin{equation}
\Sigma(R) = \int^{\infty}_{-\infty} \nu(r) dz = 2 \int^{\infty}_{R} \frac{\nu(r) r}{\sqrt{r^2-R^2}} \text{d}r.
\end{equation} 

The kinetic or pressure term on the left-hand side of Eq. \ref{PVT} is dependent only on observable quantities and describes the global average of the LOS velocity dispersion weighted by the surface luminosity. For a set of $N$ LOS velocity measurements $\{v_z\}$ spanning the entire radial extent of a galaxy, the kinetic term (after normalizing by the total number of stars) can be estimated by simply taking the mean of all squared velocities $\overline{v^2_z}$. 

There are several useful features of the projected Virial theorem, a couple of which we will mention here:-

First, as we already mentioned in the introduction, the projected Virial theorem indicates that the global energetics of the system are dependent only on the potential. The global luminosity weighted average of the LOS dispersion present in the kinetic term $K_z$ is therefore independent of the velocity anisotropy parameter $\beta$,

 All solutions to the Jeans equation automatically satisfy the projected Virial theorem so they are not two independent constraints. Rather, the projected Virial theorem gives us the information from the Jeans equation that is independent of $\beta$.

 It has been shown \citep{Wolf} that the Jeans equation only gives tight constraints on a dwarf spheroidal galaxy's total mass at the radius where the logarithmic slope of the stellar density profile $\rm{d}\ln\nu(r)/\rm{d}\ln r$ is -3, a point which often coincides with the half-light radius. At all other radii, the true mass is masked to some extent by the degeneracy with $\beta$. The projected Virial theorem is sufficient to extract the same information.

Systems with radially biased orbits have larger LOS velocity dispersions at the galactic centre where the LOS velocity component is aligned with the radial velocity component. Similarly at large radii the LOS velocity maps primarily on to the tangential velocity component which is suppressed in radially biased systems. The projected Virial theorem tells us that the luminosity weighted average of the LOS dispersion is independent of anisotropy. The dip in the LOS dispersion at large radii must therefore precisely compensate for the increased LOS dispersion at the centre of the galaxy. The projected Virial theorem forces the crossing point between the two regimes to be near to the half-light radius so that the luminosity weighted average is preserved.  

A second nice feature of the projected Virial theorem is that luminosity weighted averages automatically prioritize the richest part of the data set. For this reason it is less affected by sparse data at the galactic centre or at large radii that may bias the inference of a Jeans equation analysis when there isn't suitable flexibility in the anisotropy and density parametrization. 

In summary, the second order projected Virial theorem is often almost as useful as a full second order Jeans equation analysis.

There are two fourth order analogues of the projected Virial theorem,
\begin{equation}\label{vir4a}
\int^{\infty}_{0} \Sigma \langle v^4_z \rangle R \rm{d} R = \frac{2}{5} \int^{\infty}_{0} \nu (5-2\beta) \langle v^2_r \rangle \frac{ \rm{d}\Phi }{\rm{d}r} r^3 \rm{d}r
\end{equation}
\begin{equation}\label{vir4b}
\int^{\infty}_{0} \Sigma \langle v^4_z \rangle R^3 \rm{d} R = \frac{4}{35} \int^{\infty}_{0} \nu (7-6\beta) \langle v^2_r \rangle \frac{ \rm{d}\Phi }{\rm{d}r} r^5 \rm{d}r
\end{equation}
where, as at second order, the `pressure' terms on the left-hand side of Eqs. \ref{vir4a} and \ref{vir4b} depend only on observable quantities $\Sigma(R)$, $v_z$ and $R$.  As mentioned in the introduction, the right hand sides of these equations contain no additional parameters to those used in a second order Jeans Analysis.   Just as one may think of the projected Virial theorem at second order yielding the information contained in the second order Jeans analysis which is independent of $\beta$, the fourth order projected Virial theorem isolates information from the two fourth order Jeans equations that is independent of the anisotropy at fourth order $\beta^{\prime}$. 

The fact that Virial equations describe global averages of the velocity moments and require no radial binning of the data is even more useful at fourth order where larger sample sizes are needed for good statistics.

\subsection{Definition and Physical Properties\label{zetasec}}

In a fourth order Jeans analysis, it is a common practice (e.g \cite{Lokas05}) to normalise the fourth moment by the LOS dispersion squared to get the kurtosis. An advantage of using the kurtosis rather than the fourth moment is that there is no longer a dependence on the scale of the velocity dispersion which isolates information on the shape of the LOS velocity distribution and ensures that the same information does not count twice in any likelihood fits.

We are therefore motivated to normalise the Virial equations and to find shape parameters that are analogous to the kurtosis. While the Jeans equations describe the kurtosis at each radius the Virial equations describes global quantities. In what follows, it will be useful to define the following notation for the weighted average for various quantities:-
\begin{equation}\label{Intform}
X_\star \equiv \frac{1}{N_{\rm{tot}}} \int^\infty_0 X(R) \Sigma(R) R \;\rm{d}R. \end{equation}
which represents the normalized global average of some quantity $X(R)$ weighted by the expected number of stars at each projected radius $R$. 
  
First, it is convenient to renormalize the Virial equations by the total number of stars
\begin{equation}
N_{\rm{tot}} = \int^{\infty}_{0} \Sigma(R) R \rm{d}R
\end{equation}
so that the kinetic term on the left-hand side of the Virial equations is equivalent to simply taking the global mean of $v^4_z$ and $v^4_z R^2$ data. Dividing through by the averaged LOS dispersion then gives us the Virial shape parameters,
\begin{equation}\label{zetaA}
\zeta_A = \frac{\langle v^4_z \rangle_\star}{\langle v^2_z \rangle^2_\star} = \frac{9 N_{\rm{tot}} }{10} \frac{\int^{\infty}_{0} \nu (5-2\beta) \langle v^2_r \rangle \frac{ \rm{d}\Phi }{\rm{d}r} r^3 \rm{d}r}{ \left( \int^{\infty}_{0} \nu \frac{ \rm{d}\Phi }{\rm{d}r} r^3 \rm{d}r \right)^2}
\end{equation}
\begin{equation}\label{zetaB}
 \zeta_B = \frac{ \left( \langle v^4_z \rangle R^2 \right)_\star}{\langle v^2_z \rangle_\star^2 R^2_\star} = \frac{9 N^2_{\rm{tot}} }{35} \frac{\int^{\infty}_{0} \nu (7-6\beta) \langle v^2_r \rangle \frac{ \rm{d}\Phi }{\rm{d}r} r^5 \rm{d}r}{\left( \int^{\infty}_{0} \nu \frac{ \rm{d}\Phi }{\rm{d}r} r^3 \rm{d}r \right)^2 \int^{\infty}_{0} \Sigma(R) R^3 \rm{d}R}.
\end{equation}
The variable $\zeta_A$ broadly translates to the luminosity-weighted average of the LOS velocity kurtosis, i.e a value of $\zeta_A<3$ indicates that on average the LOS velocity distribution is more flat topped than a Gaussian distribution in the most luminous regions of the galaxy. From Eqs. \ref{zetaA} and \ref{zetaB} we see that by normalizing the fourth moment with the LOS dispersion squared we effectively eliminate any global scaling of the system's total mass.  If the density of dark matter dominates the mass density of the stars (as in most dwarf spheroidal galaxies) then this makes the $\zeta$ variables independent of the dark matter scaling density. 

These two parameters $\zeta_A$ and $\zeta_B$ have several interesting properties, including two key properties pertinent to this work:-
\begin{enumerate}
\item
In the case where the anisotropy parameter $\beta$ is constant at all radii, for a fixed tracer density $\nu(r)$ and dark matter density $\rho(r)$ numerical tests show that $\zeta_A(\beta)$ is a monotonically {\it increasing} function of $\beta$ while $\zeta_B(\beta)$ is a monotonically {\it decreasing} function of $\beta$. We can illustrate this with a simple argument. We note that increasing $\beta$ will inflate the radial velocity dispersion $\langle v^2_r \rangle$ but not the total velocity dispersion $\langle v^2 \rangle = (3-2\beta) \langle v^2_r \rangle$, which the Virial theorem tells us is dependent only on the gravitational potential. If we rewrite the $\beta$ dependence of $\zeta_A$ as $(5-2\beta)\langle v^2_r \rangle = \langle v^2 \rangle + 2 \langle v^2_r \rangle$ and $\zeta_B$ as $(7-6\beta)\langle v^2_r \rangle = 3 \langle v^2 \rangle - 2 \langle v^2_r \rangle$ then it is clear that $\zeta_A$ will increase with $\beta$ and $\zeta_B$ will decrease. If we recall that $\zeta_A$ may be interpreted as the luminosity weighted average of the LOS kurtosis, this result is consistent with the observation \citep{dejonghe87,Gerhard93,Lokas05} that systems with tangential biased orbits ($\beta<0$) have declining kurtosis profiles and that conversely systems with radially biased orbits ($\beta>0$) have rising kurtosis profiles.
\item
Secondly, for fixed anisotropy $\beta$, both $\zeta$ parameters increase for more concentrated stellar populations. Specifically we find that the $\zeta$ parameters increase if the scale radius of the stellar population $r_\star$ is made smaller relative to the scale radius $r_{s}$ of the DM halo. Dark matter haloes with shallow, cored central density slopes have larger $\zeta_A$ than cusped profiles. As $\zeta_A$ weights the velocity moments most heavily where there are most stars then if the stars are deeply embedded into the dark matter halo we can expect the difference in $\zeta_A$ between cored and cusped profiles to be amplified because most of the weight is applied in the inner regions below the DM scale radius where the density profiles are most different. By contrast, $\zeta_B$ has an additional $R^2$ that places a greater weight to stars at large projected radii $R$ and it is therefore more sensitive to the density profile in the outer parts. Even if the stars are deeply embedded into the dark matter halo, differences between the outer slopes of dark matter density profiles can therefore mask differences between cores and cusps present in $\zeta_B$.
\end{enumerate}

\subsection{Statistical Properties}\label{zetastat}
Next we examine the statistical properties of the $\zeta$ parameters to see how well they can be recovered from a limited sample of stellar positions and LOS velocities. From the left-hand side of Eqs. \ref{zetaA} and \ref{zetaB} simple estimators for the $\zeta$ parameters are,
\begin{equation}\label{zetaestA}
\widehat{\zeta}_A = N_{\rm{s}} \frac{\sum^{N_{\rm{s}}}_i v^4_{z,i} }{\left(\sum^{N_{\rm{s}}}_i v^2_{z,i} \right)^2 }
\end{equation}
\begin{equation}\label{zetaestB}
\widehat{\zeta}_B = N^2_{\rm{s}} \frac{\sum^{N_{\rm{s}}}_i v^4_{z,i} R^2_i }{\left(\sum^{N_{\rm{s}}}_i v^2_{z,i} \right)^2 \sum^{N_{\rm{s}}}_i R^2_i }.
\end{equation}
which can be read directly from the full set of $N_{\rm{s}}$ LOS velocities and projected radii in either simulated or real data. Ultimately, we want to add the measured values of $\widehat{\zeta}$ to the LOS dispersion measurements in a joint likelihood analysis. To do this we must first estimate the variance and bias of the $\widehat{\zeta}$ error distribution with numerical tests.
  
\begin{table}
\caption{Grid of simulated dSph model parameters from the \textit{Gaia} challenge test suite presented in \protect\cite{gaiamodel}. Both the tracer stars (with extra $\star$ subscript) and the dark matter have density profiles parametrized by the Zhao model (see Eq. \protect\ref{zhao}). For each of the 16 combinations below the test suite has two variants for the anisotropy parameter: isotropy $\beta=0$ and Osipkov-Merritt anisotropy described by Eq. \protect\ref{bvh} with $\beta_0 = 0,\beta_\infty=1$ and $r_\beta=r_\star$.}
\centering
\begin{tabular}{ L{2.2cm} L{2cm} L{3.5cm} }
\hline
Profile & Parameter & Values Considered \\
\hline
Tracer Density & & \\
$\nu(r)$ & $r_\star$[kpc] & 0.1,\;0.25,\;0.5,\;1  \\
 &  $\alpha_\star$ & 2 \\
 &  $\beta_\star$ & 5 \\
 &  $\gamma_\star$ & 0.1,\;1 \\
 DM density & & \\
 $\rho(r)$ &  $\rho_{0}$ [$M_\odot$pc$^{-3}$] & 0.4 ($\gamma=0$) \\
 & & 0.064($\gamma=1$) \\  
 &  $r_{\rm{s}}$[kpc] & 1  \\
 &  $\alpha$ & 1 \\
 &  $\beta$ & 3 \\
 &  $\gamma$ & 0,\;1 \\
\hline
\end{tabular}
\label{gaiaparams}
\end{table} 

The \textit{Gaia} challenge (\cite{gaiamodel}) test suite consists of simulated data from spherical models of dwarf spheroidals that have self-consistent distribution functions spanning a variety of tracer densities $\nu(r)$, anisotropy parameters $\beta(r)$ and dark matter density profiles $\rho(r)$. In the \textit{Gaia} challenge models, $\nu(r)$ and $\rho(r)$ are parametrized by $\alpha,\beta,\gamma$ (sometimes referred to as Zhao) profiles,
\begin{equation}
\nu(r)=\frac{\nu_0}{\left(\frac{r}{r_\star}\right)^{\gamma_\star}\left[1+\left(\frac{r}{r_\star}\right)^{\alpha_\star}\right]^{(\beta_\star-\gamma_\star)/\alpha_\star}}.
\label{tracezhao}
\end{equation}
\begin{equation}
\rho(r)=\frac{\rho_0}{\left(\frac{r}{r_s}\right)^\gamma\left[1+\left(\frac{r}{r_s}\right)^\alpha\right]^{(\beta-\gamma)/\alpha}}.
\label{zhao}
\end{equation}
 This density profile has five parameters and the inner slope $\gamma$, the outer slope $\beta$ and the rate (in $r$) of transition between the two $\alpha$ can be varied. The parameters $\rho_0$ ($\nu_0$) and $r_s$ ($r_\star$) give the normalization and scale radius of the dark matter (stellar) density profile respectively. The velocity anisotropy $\beta$ of the \textit{Gaia} challenge models may be parametrized by \citep{baes07},
\begin{equation}\label{bvh}
\beta(r) = (\beta_{\infty}-\beta_0) \frac{r^2}{r^{2}_{\beta}+r^2} + \beta_0
\end{equation}
which describes a generic radial profile that increases/decreases monotonically between asymptotic values $\beta_0$ at $r=0$ and $\beta_{\infty}$ at $r = \infty$ about a scale radius $r_\beta$.

A grid of model parameters for the 32 unique \textit{Gaia} challenge models is shown in Table. \ref{gaiaparams} and we refer the reader to \cite{penarrubia} for a plot of the LOS dispersion profiles. For each model we can use the model parameters in Table \ref{gaiaparams} to calculate the 'true' value of the Virial shape parameters $\zeta$ which will enable us to compute the bias $\widehat{\zeta}-\zeta$.

A large database of over $10^5$ simulated stars is provided for each of the 32 models from which we can randomly draw many galaxy sized samples. Because we are ultimately interested in looking at Sculptor, we fix the total number of stars in each sample to be the same as in the Sculptor data-set  $N_s=1350$. For each model in the \textit{Gaia} challenge test suite we can calculate error distributions for $\widehat{\zeta}$ by 1) drawing many Monte Carlo samplings of $N_s=1350$ stars from the full set of $10^5$, 2) calculating $\widehat{\zeta}$  for each galaxy sized sample and 3) making a histogram of the resulting list of $\widehat{\zeta}$ measurements.

The \textit{Gaia} challenge test suite also provides simulations of systematics such as experimental velocity errors, binary stars and Milky Way interlopers that the user is free to turn off or on. By introducing the systematics one at a time we can see what impact they have on the bias. A detailed discussion of how various systematics affect the bias and how they are modelled is provided in the Appendix. 

\begin{figure}
	\centering
		\includegraphics[width=9cm]{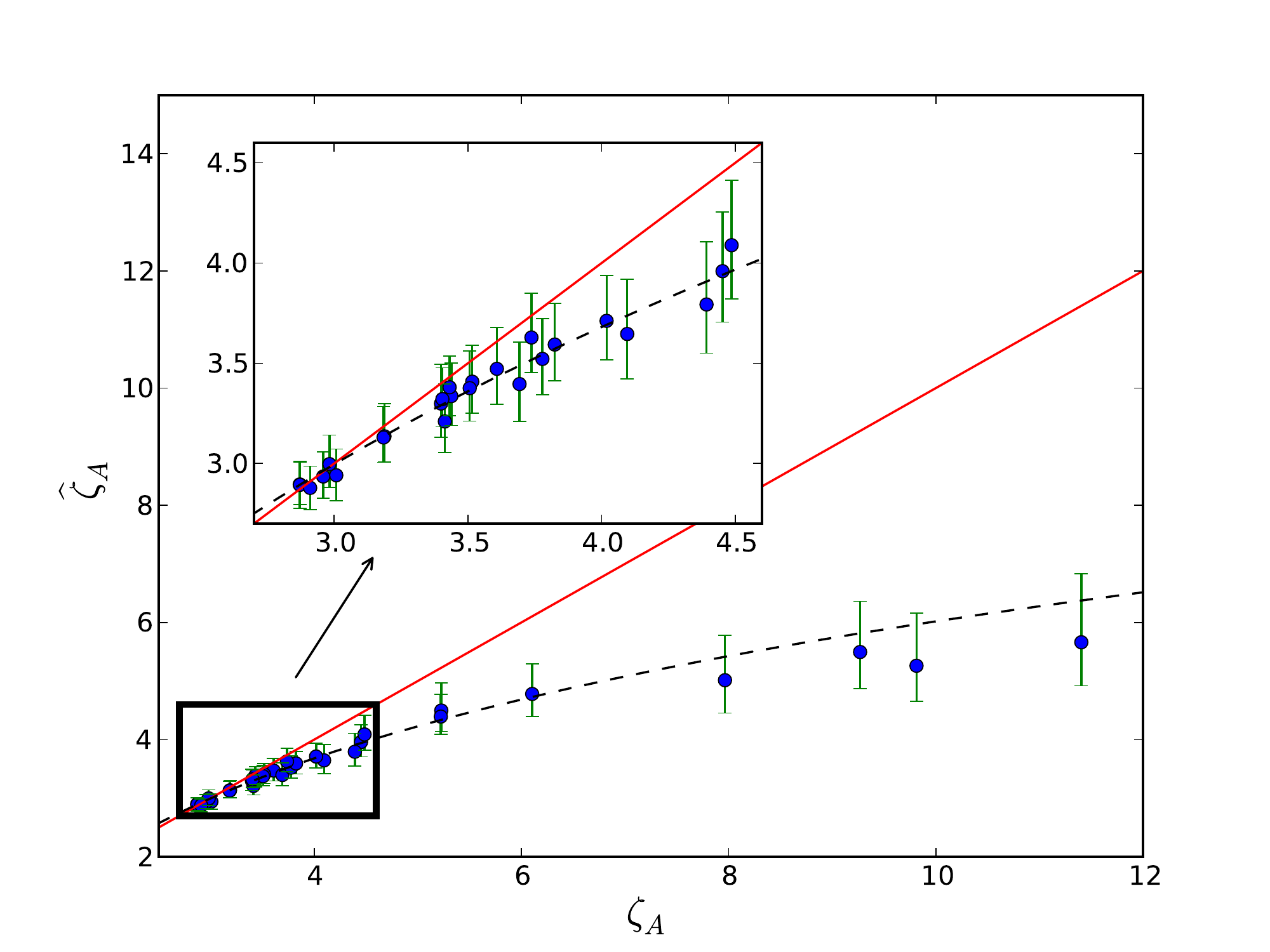}
       	\caption{Numerical probability distributions of the estimator $\widehat{\zeta}_A$ were calculated from Monte Carlo samplings of all 32 \textit{Gaia} challenge models described in Table \ref{gaiaparams}. Central data points and error bars show the median value and the boundaries enclosing the central $67\%$ of each distribution. The abscissa of each point is the `true' value of $\zeta_A$ corresponding to the model parameters in Table \ref{gaiaparams} and Eq. \ref{zetaA}. The red line therefore shows the prediction for the estimator $\zeta_A$ if it is unbiased. The dashed black line shows a fit to the bias with a simple power law curve.}
	\label{gaiaepA}
\end{figure}

\begin{figure}
	\centering
		\includegraphics[width=9cm]{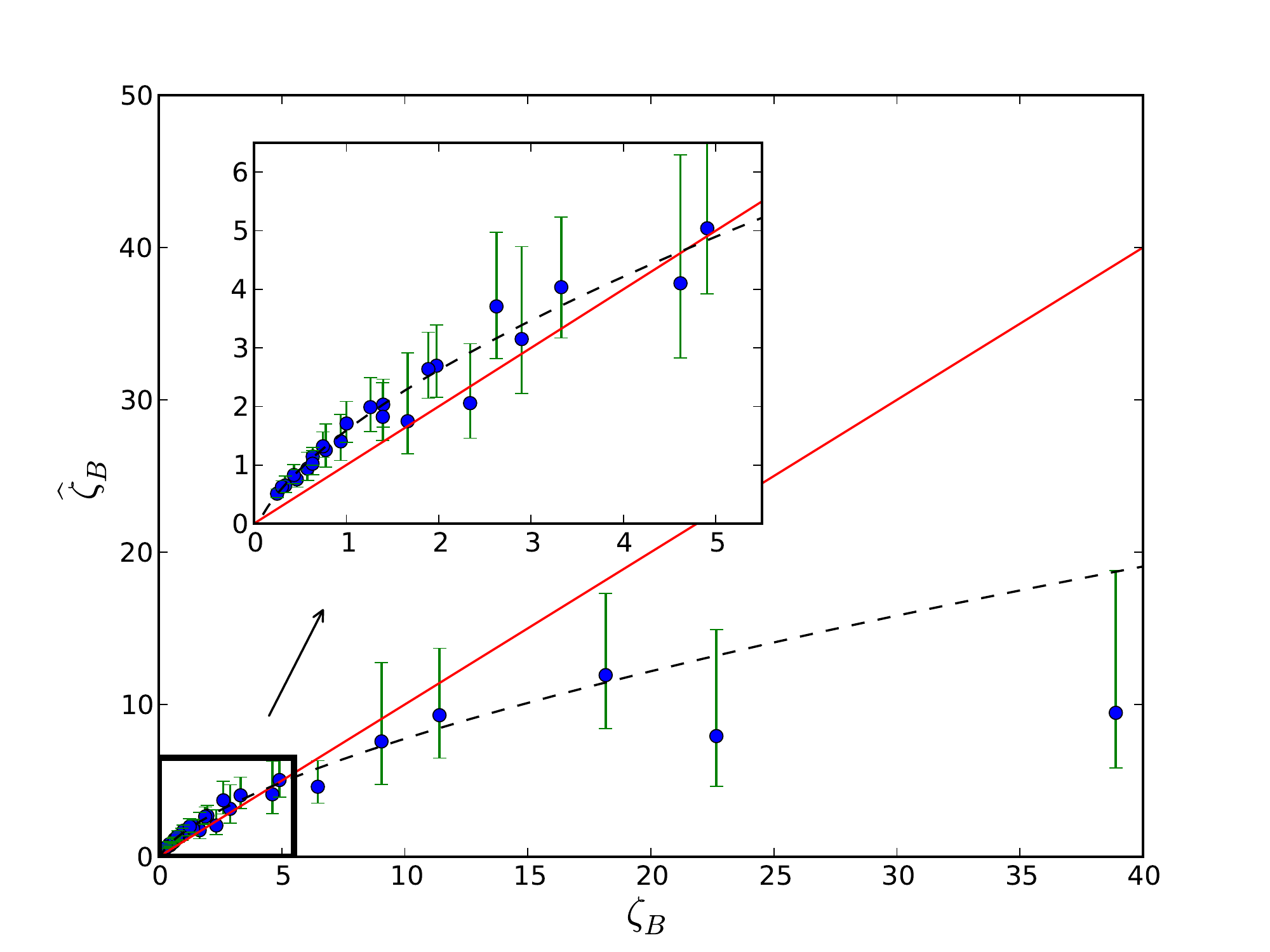}
       	\caption{Numerical probability distributions of $\widehat{\zeta}_B$. Key as per Fig. \ref{gaiaepA}.}
	\label{gaiaepB}
\end{figure}
For all 32 \textit{Gaia} challenge models, Figs. \ref{gaiaepA} and \ref{gaiaepB} show the net contribution from all three sources of bias that we consider in our final analysis:- experimental errors, binary motion for a fraction $f_b=0.6$ of stars and the interloper removal scheme.

In Fig. \ref{gaiaepA}, we see that when the true value $\zeta_A$ is large then the net bias pushes the estimator $\widehat{\zeta}_A$ back towards the Gaussian value of 3. This is what we might expect given a convolution with experimental errors that are approximately Gaussian distributed. Surprisingly, despite the binary velocity distribution bearing no resemblance to the Gaussian distribution (see Appendix), we found that binary stars also pushed the bias downwards. The asymmetric error bars on the right hand side of Fig. \ref{gaiaepA} also show that when $\zeta_A >> 3$ the $\widehat{\zeta}_A$ distributions become increasingly skewed and less well approximated by a Gaussian fit. At $\zeta_A = 3$, we see that the bias vanishes and that the error bars are symmetric. In other words, we can naively say that if the intrinsic LOS velocity distribution at all radii is approximately Gaussian then $\zeta_A$ is not significantly biased. By choosing normalized shape parameters rather than the fourth moments, the bias from Gaussian experimental error distributions naturally cancels out because the convolution of two Gaussians is still Gaussian in shape.   

Despite the fact that the 32 models described in Table \ref{gaiaparams} cover a diverse range of parameters both the bias and the variance of the $\widehat{\zeta}_A$ distributions correlate strongly with $\zeta_A$. For this reason we approximate the bias and variance of the $\widehat{\zeta}_A$ distribution in Sculptor by taking the observed value of $\widehat{\zeta}_A$ and interpolating the bias and variance curves shown in Fig. \ref{gaiaepA} for the \textit{Gaia} challenge models.                

Our numerical tests show that there is a much larger (model to model) scatter in the bias and variance of $\zeta_B$. In other words, the error in $\widehat{\zeta}_B$ that we measure is more sensitive to the anisotropy and density profile. While we found that $\zeta_A$ is not strongly biased by the interloper removal scheme, the additional $R^2$ weighting in $\zeta_B$ makes it much more sensitive to removing stars at large radii that are deemed likely to be Milky Way contaminants. We found that the interloper removal algorithm dominates the bias in $\zeta_B$ and in the Appendix we discuss how this gives rise to the more complicated bias curve that we observe in Fig. \ref{gaiaepB}.

For both $\zeta$ parameters, the best fitting lines and the error on the recovery of $\widehat{\zeta}_i$ from the actual value of $\zeta_i$ is obtained by using the Metropolis-Hastings algorithm to perform many fits to the data with a function of the form  $\widehat{\zeta}_i=a\zeta_i^b+c$. In this way we obtain a list of $a,b$ and $c$ which gives rise to a cumulative distribution function in $\zeta$ for each value of $\widehat{\zeta}$.

\section{Illustrative Analysis (fits with constant $\beta$)}\label{fixbetsec}

In this section we perform a standard Jeans analysis fitting to velocity dispersion data only.  We will then check to see which regions of parameter space favoured by the standard Jeans analysis are also compatible with the measured value of $\widehat{\zeta}$ in order to investigate how the new parameter constrains the fit.  

We will use rich data from the dwarf spheroidal galaxy Sculptor as there is evidence for a kurtosis which is different from Gaussian at a statistically significant level. We therefore hope that this will make it a good object to see the power of higher order techniques.
For the time being we will restrict our fit to consider only solutions of the Jeans equations with constant velocity anisotropy $\beta$, a fixed Plummer profile (Eq. \ref{tracezhao} with $\alpha_\star=2$, $\beta_\star=5$ and $\gamma_\star=0$) for the stars and simple two parameter density profiles so that we can see more clearly how $\widehat{\zeta}$ can distinguish between different solutions. Later sections will go on to consider the more realistic situation of generalized non-constant beta and will add the $\zeta$ parameters to the likelihood function in the initial fit rather than as a filter applied after the initial second order Jeans analysis.

\subsection{Sculptor: photometry and kinematics}
The Sculptor data set that is used throughout this work comprises of the RA-Dec coordinates and heliocentric rest frame velocities from the Magellan survey published in \cite{walkdat}. We adopt a distance to Sculptor of $d=79$kpc and central RA-Dec coordinates of $\alpha_C=01^h00^m09^s$ and $\delta_C= -33^{o} 42.5^{\prime}$ from \cite{mateo} to estimate the projected radius $R^2=x^2+y^2$ (recall that we choose $z$ as the LOS direction) of each star from its RA-Dec coordinates. To subtract the component along the LOS of the dwarf spheroidal's bulk velocity relative to the Sun we transformed all of the velocities into the dwarf spheroidal rest frame as described in \cite{walkerrot} and removed the mild velocity gradients. As we will discuss in the Appendix, this procedure had no significant effect on our results. Finally we removed stars with a probability of membership less than 0.95 according to the interloper removal procedure described in \cite{walkermetal} that accompanies the kinematic data set. We discuss the reasons for this choice in the Appendix and compare our results with those obtained with other interloper removal schemes.

Throughout this section we fix the tracer density to be the best fitting \citep{irwin} Plummer profile (with $r_\star = 260$pc and $\nu_0$ normalized to a give a total luminosity of $L_{\rm{tot}} = 1.4 \times 10^6 L_{\odot}$). In our benchmark model we choose a mass-luminosity ratio of $\Upsilon_\star=1M_\odot/L_\odot$ (as in \cite{strigkurt,breddels}) for the stellar population and split the velocity data equally into $N_b=26$ radial bins when calculating the LOS dispersions. We tested each of these assumptions and the implications of our choices are discussed at length in \S4.

\subsection{Consistency check for the standard second order jeans analysis}
 In a typical Jeans analysis we are presented with a data set $d = \{R_i,v_{z,i}\}$ of projected radii $R_i$ and LOS velocity measurements $v_{z,i}$. The velocity data are then split into $N_b$ radial bins such that there are a set of $N_b$ dispersion (i.e variance) measurements $S_{2,j}$ and $N_b$ bin radii $R_j$. The tracer density $\nu(r)$, anisotropy $\beta(r)$ and DM density profile are all parametrized and form a parameter space $P$. For any individual set of parameters $p \in P$, the Jeans equation \citep{binney},
\begin{equation}\label{jeans}
\frac{d\nu \langle v^{2}_{r}\rangle}{dr} + \frac{2\beta}{r}\nu \langle v^{2}_{r}\rangle +\nu \frac{d \Phi}{dr} = 0.
\end{equation}
is used to calculate $\langle v^2_r \rangle (p)$ which may then be used in turn to calculate the LOS velocity dispersion $\langle v^2_z \rangle(p)$ via \citep{binney},
\begin{equation}\label{LOSsecond}
\Sigma \langle v^{2}_{z} \rangle (R) = 2\int^{\infty}_{R} (1-\beta\frac{R^2}{r^2}) \frac{\nu \langle v^{2}_{r}\rangle r}{\sqrt{r^2-R^2}}dr. 
\end{equation}
Under the assumption that the variance measurements $S_{2,j}$ are Gaussian distributed then the likelihood function (see e.g \cite{universal,martinez}),
\begin{equation}\label{likedisp}
\mathcal{L}(d|p) = \prod^{N_b}_{j} \frac{1}{\sqrt{2 \pi \rm{Var}[S_{2,j}]}} \exp \left\{-\frac{ \left[S_{2,j} - \langle v^2_z \rangle (R_j|p) - b_j\right]^2}{2 \rm{Var}[S_{2,j}]}\right\}
\end{equation}
can be used to derive the posterior distributions $\mathcal{P}(p|d)$ of each parameter with Bayesian inference methods such as Monte Carlo Markov Chains or nested sampling algorithms.  Here $b_j=\rm{Mean}[S_{2,j}]-\langle v^2_z \rangle(R_j|p)$ is the bias on the mean of the $S_{2,j}$ distribution from the addition of experimental errors and binary stars that artificially inflate the LOS velocity dispersion. We use bootstrap sampling of the \textit{Gaia} challenge data to estimate the bias and the variance $\rm{Var}[S_{2,j}]$ of the $S_{2,j}$ error distribution. To generate samples from the posterior distributions $\mathcal{P}(p|d)$ we use the nested sampling code MULTINEST \citep{multinest}.

As discussed previously, the parameter space $P$ that is used to find the LOS dispersion $\langle v^2_z \rangle$ from the Jeans equation is sufficient to calculate the $\zeta$ variables via the right-hand side of Eqs. \ref{zetaA} and \ref{zetaB}. For each parameter set $p$ in the posterior sample that fits the LOS velocity dispersion we can calculate the $\zeta$ parameters and compare them to the estimators $\widehat{\zeta}$ from the data.  This gives us a free consistency check between the width and the shape of the LOS velocity distribution that can be used to break the mass-anisotropy degeneracy.

\begin{figure*}
	\centering
		\includegraphics[width=17cm]{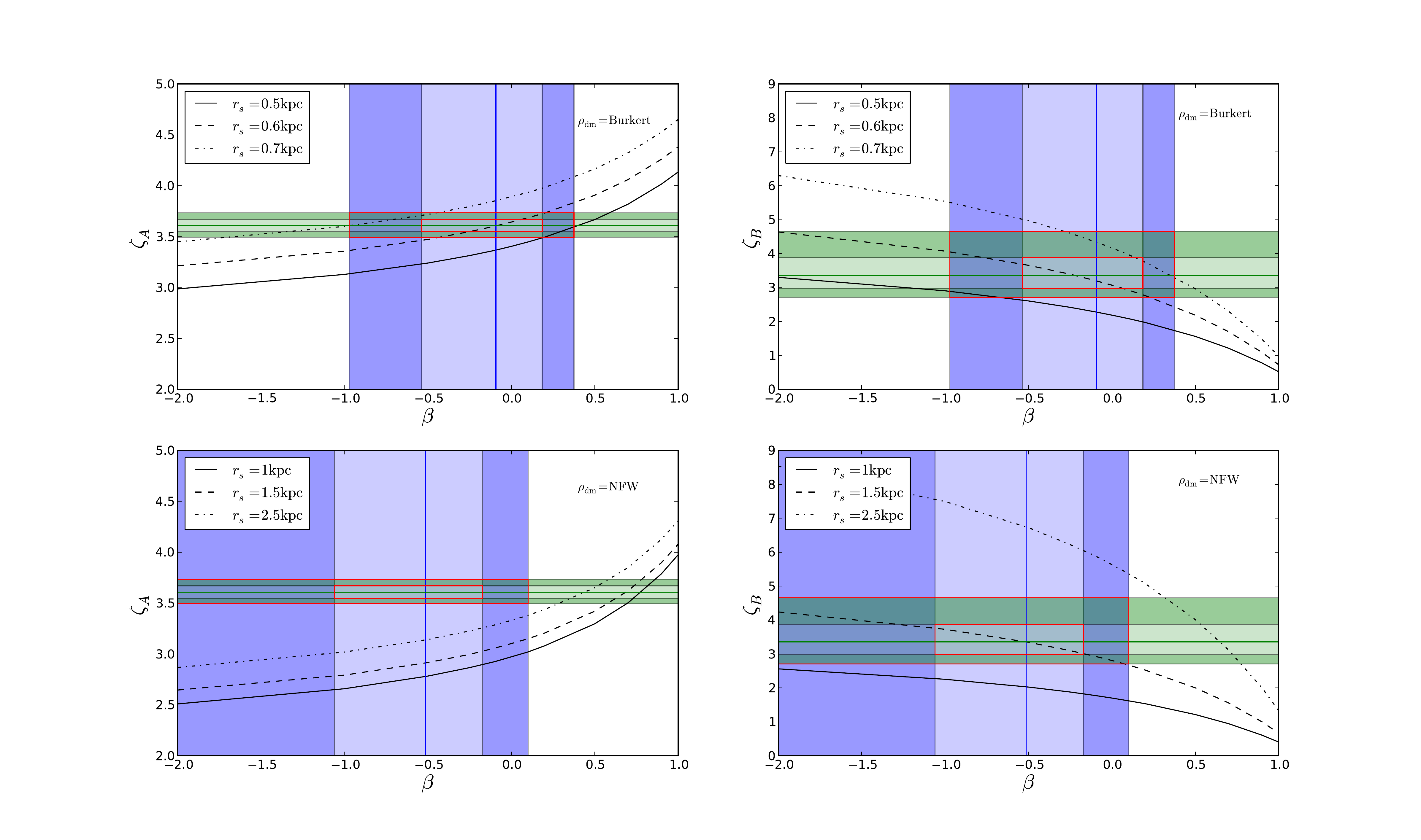}
       	\caption{Virial theorem consistency check. Blue regions (vertical stripes) show the 67\% and 95\% for the best $\chi^2$ probability for each value of the velocity anisotropy parameter $\beta$ in a Jeans Equation fit to the LOS velocity dispersion of Sculptor. Green regions (horizontal stripes) are estimates of the 1$\sigma$ and 2$\sigma$ likelihood contours for $\zeta_A$ given the measurement of $\widehat{\zeta}_A = 3.43$ from the Sculptor velocity set and the bias and error bars from the best fitting interpolation of the \textit{Gaia} challenge mock data. Black lines show $\zeta(\beta)$ as derived from Eqs. \ref{zetaA} and \ref{zetaB} with a Plummer profile and self-gravity with $\Upsilon_\star=1$.}
	\label{sculptconstep}
\end{figure*}

\subsection{Results for Sculptor with constant $\beta$}
In Fig. \ref{sculptconstep} we perform the consistency check on the Sculptor data set. In this illustrative example we fix the anisotropy to be constant as a function of radius. These assumptions will be relaxed later. In the upper panel we study a cored Burkert density profile which takes the following form
\begin{equation}
\rho(r)=\frac{\rho_0r_s^3}{\left(r_s+r\right)\left(r_s^2+r^2\right)}
\label{burkeq}
\end{equation}
where $r_s$ and $\rho_0$ are the two free parameters of the model, the scale radius and the characteristic density. We compare it with the cusped Navarro-Frenk-White (NFW) profile in the bottom panel, the NFW profile has the following form
\begin{equation}
\rho(r)=\frac{\rho_0r_s^3}{r\left(r^2+r_s^2\right)}
\label{nfweq}
\end{equation}
The NFW profile corresponds to Eq. \ref{zhao} with $\alpha=1$, $\beta=3$ and $\gamma=1$ but we note that the Burkert is not nested within the Zhao parametrization in this way.
We varied $r_s$, $\rho_0$ and the constant anisotropy parameter $\beta$ in order to find the best $\chi^2$ fits to the LOS dispersion data in a standard second order Jeans analysis. For the standard Jeans analysis presented in this section the relevant $\chi^2(p)$ for a parameter set $p=\{\beta,\rho_0,r_s\}$ corresponding to Eq. \ref{likedisp} is
\begin{equation}\label{chi2jeans}
\chi^2(p) = \sum^{N_b}_{j} \frac{ \left[S_{2,j} - \langle v^2_z \rangle (R_j|p) - b_j\right]^2}{\rm{Var}[S_{2,j}]}
\end{equation}
and we may derive a p-value by considering a $\chi^2$ distribution with the relevant degrees of freedom. Due to the mass-anisotropy degeneracy, the standard Jeans analysis will often have a likelihood surface that isn't sharply peaked which makes the choice of priors important. Since we are not really certain of the priors on the various parameters coming from astrophysics we use the likelihood to determine which models better fit the data rather than the posterior distributions. 

Vertical blue stripes in Fig. \ref{sculptconstep} show the values of $\beta$ for which the p-value falls to $0.167$ and $0.05$.

Let us now introduce the $\zeta$ variables for the Sculptor data. Using Eqs. \ref{zetaestA} and \ref{zetaestB} we found $\widehat{\zeta}_A = 3.43$ and $\widehat{\zeta}_B = 3.69$. Green likelihood regions in Fig. \ref{sculptconstep} show the 1$\sigma$ and 2$\sigma$ confidence regions for $\zeta$ that are discussed in \S\ref{zetastat}. We re-emphasize that these regions are derived purely from the data and therefore remain fixed in both panels. The large observed value of $\zeta_A \sim 3.5$ indicates that Sculptor has a more peaked LOS velocity distribution than a Gaussian $\zeta_A=3$ which is consistent with an analysis \cite{Amorisco} made of the same data set with Gauss-Hermite moments.

Red squares in Fig. \ref{sculptconstep} show the regions where the observed value of $\widehat{\zeta}$ is consistent with the anisotropy $\beta$ derived from fits to the LOS velocity dispersion. For two-parameter Burkert and NFW density profiles the remaining degrees of freedom are the scale density $\rho_s$ and scale radius $r_s$. By design, the $\zeta$ variables are normalized to remove the dependence on the scale density if the stellar self-gravity is negligible. We found that in the case where the tracer mass-luminosity ratio $\Upsilon_\star =1$ then $\zeta$ is almost completely insensitive to the scale density. Even if this wan't the case, if the scale radius is fixed then the projected Virial theorem places tight constraints on the scale density (which we use to adopt $\rho_0$ in the figure). The scale radius of the dark matter halo is therefore effectively the only remaining degree of freedom in Fig. \ref{sculptconstep}. Black lines show $\zeta_A$ and $\zeta_B$ for different choices of the dark matter scale radius. For a model to be consistent with the Sculptor data set, the black lines must pass through the red squares.

 The most striking feature of the figure is that $\zeta_A$ curves for the NFW density profiles only pass through the 2 sigma region when the NFW scale radius is large with $r_{s} \sim 2.5\rm{kpc}$ being almost ten times larger than the half-light radius of the stellar population. haloes with smaller scale radii have velocity distributions that are too flat-topped to describe the Sculptor velocity data. Interestingly, \cite{evansvirial} find that the energetics of two stellar sub populations in Sculptor are fit excellently by cored DM haloes and can only \emph{approach}\footnote{We assume that NFW profiles with smaller self-gravity than those  presented in fig. 1 of \cite{evansvirial} with $\Upsilon_\star=8$ provide better fits in line with the results presented for cored density profiles.} a $2\sigma$ level of agreement in a shared NFW profile if the scale radius is large. 

\begin{figure}
	\centering
		\includegraphics[width=9cm]{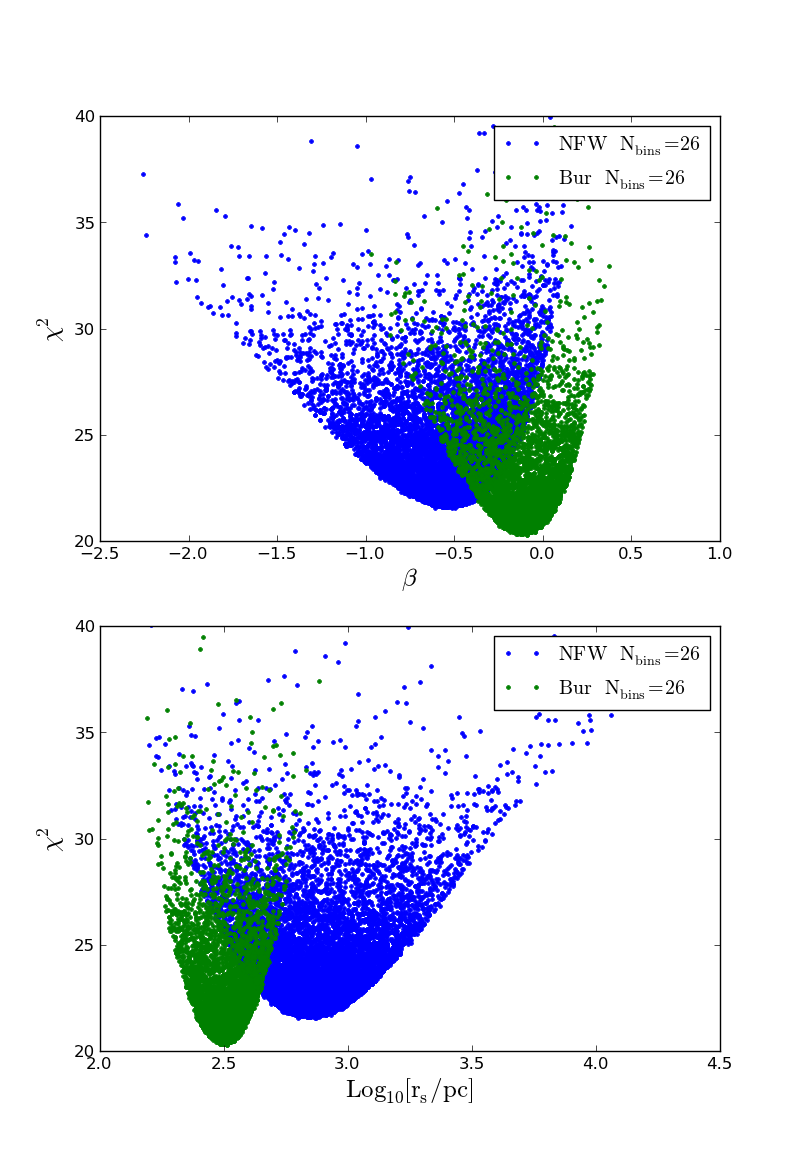}
       	\caption{For a standard second order Jeans fit to Sculptors LOS velocity dispersion we show the $\chi^2$ (Eq. \ref{chi2jeans}) goodness of fit as a function of the anisotropy parameter $\beta$ and the scale radius of the DM halo $r_s$. The fit has 22 degrees of freedom.}
	\label{constbetchi}
\end{figure}

From Fig. \ref{constbetchi} we see that NFW fits with large scale radii of $r_s>2.5$kpc ($\rm{Log}_{10} r_s > 3.4$pc) begin to incur a significant $\chi^2$ penalty when fitting to the LOS dispersion data if $\beta$ is held constant. As we discuss in the Appendix, fits to the dispersion alone are sensitive to the choice of binning and selecting fewer radial bins reduces the tension with large scale radii.

By contrast, Burkert models with much smaller scale radii $r_{s}\sim 0.6\rm{kpc}$ are perfectly consistent with the peaked velocity distribution implied by $\zeta>3$ and can simultaneously fit both $\zeta$ parameters. These scale radii are consistent with a scaling relation for cored dark matter haloes in spiral galaxies (see fig. 2 of \cite{salucciwalker}).
 
We note however that this best fitting solution isn't a good fit to the LOS dispersion data. In Fig. \ref{constbetchi} we see that Burkert fits to the dispersion in our Jeans analysis prefer lower scale radii (Log $r_s \sim 2.5$pc). The best fitting value to the $\zeta$ parameters, Log$_{10}r_s \sim 2.8$, sits relatively high on the $\chi^2$ curve. The statistical significance of this is discussed later in the generalized analysis.

\subsection{General points on interpreting the $\zeta$ parameters}\label{zinterp}

Our aim in this section is to use the results above to illustrate more general comments on how the properties of the $\zeta$ parameters outlined in \S\ref{zetasec} can help to distinguish between different DM density profiles. In Fig. \ref{sculptconstep} we can confirm visually that (i) for a fixed density profile (black curve) increasing $\beta$ will increase the ratio $\zeta_A/\zeta_B$, (ii) more concentrated DM haloes (lower $r_s$) have lower values of both $\zeta$ parameters  and (iii)  if $\beta$ and $r_s$ is fixed then cored DM density profiles have larger values of both $\zeta$ parameters than cusped ones.   

From the blue regions in Fig. \ref{sculptconstep} we see that the best fitting NFW profiles have a greater degree of tangential anisotropy than the Burkert profiles. To fit the same flattish LOS dispersion profile, the cusped NFW model (with a larger gravitational potential at the galactic centre) must have anisotropy that is tuned to reduce the LOS velocity dispersion in these regions relative to the cored Burkert profile. This is achieved by decreasing the anisotropy parameter $\beta$ for the NFW profile so that in central regions, where the LOS velocity is dominated by the radial velocity component, the LOS velocity dispersion is suppressed. The same result is illustrated in the left-hand panel of fig. 1 in \cite{Wolf} where the authors show that the best fitting radially anisotropic models $\beta>0$ have small interior masses relative to the tangentially anisotropic ones.

 As cusped profiles will often have more tangential anisotropy than cored fits to the same dispersion data we see that only property (ii) listed above can be used to boost $\zeta_A$ relative to cored profiles. In other words, cusped fits to $\zeta_A$ data will often have larger scale radii than cored fits as we see in Fig. \ref{constbetchi}. We confirm this in analyses with generalized anisotropy in later sections. Because the velocity distribution in Sculptor is more peaked than Gaussian this difference in scale radii must be particularly large to match $\zeta_A>3$. The degeneracy in $\zeta_A$ between the inner slope and the scale radius can then be broken by $\zeta_B$ and in Fig. \ref{sculptconstep} we see that the cusped model with a large scale radius cannot simultaneously fit both $\zeta$ parameters.          


To summarize, in this section we have considered the overly simplified situation where the velocity anisotropy parameter $\beta=$ constant since in this case there is a one-to-one mapping between velocity anisotropy and the $\zeta$ parameters.  Within this idealized setting we demonstrated the power of $\zeta_A$ and $\zeta_B$ in distinguishing between cuspy and cored profiles since cuspy models give rise to values of $\zeta_A$ that are incompatible with the data.

In the next section we will examine what happens when we apply the $\zeta$ parameters to the more realistic situation where $\beta$ is a function of radius.

\section{Realistic Analysis (fits with generalized $\beta$)\label{genbetsec}}

We now outline a more realistic approach to evaluate which density profiles the Sculptor data set is most consistent with. Throughout this section the assumption of constant $\beta$ is relaxed and without strong prior intuition for the anisotropy parameter $\beta(r)$ we use generalized parametric form introduced in Eq. \ref{bvh}.   

In this section, we will also widen the range of density profiles by using the more general $\alpha,\beta,\gamma$ density profile. As well as the Burkert  profile (Eq. \ref{burkeq}) which cannot be described by the $\alpha,\beta,\gamma$ parametrization we will consider the NFW profile (Eq. \ref{nfweq}) and other nested models where $\alpha,\beta,\gamma$ are fixed. Finally, we will consider the case when all three parameters are free to vary. 

We perform a conventional Jeans equation analysis of Sculptor's LOS velocity data for a generic space $P$ of parameters for the anisotropy $\beta(r)$ and dark matter density $\rho(r)$.  We do this with and without the fits for the value of $\zeta_A$ and $\zeta_B$ for the same set of parameters $P$. When we do include one or both of the $\zeta$ parameters in our MULTINEST runs we multiply the likelihood function for the dispersion data (Eq. \ref{likedisp}) by the likelihood functions $\mathcal{P}(\widehat{\zeta}|\zeta)$ that we obtained from the analysis of the \textit{Gaia} challenge data outlined in \S\ref{zetasec}. 

As the numerically determined likelihood functions $\mathcal{P}(\widehat{\zeta}|\zeta)$ are not pure Gaussian functions we use a pseudo-$\chi^2$ statistic,
\begin{equation}\label{pseudochi}
\chi^2 = -2(\rm{Ln}\mathcal{L}-\rm{Ln}\mathcal{L}_{\rm{max}})
\end{equation}which is exactly equivalent to the conventional $\chi^2$ statistic if the likelihood $\mathcal{L}$ is a normalized Gaussian and peaked at $\mathcal{L}_{\rm{max}}$.  We found that $\mathcal{P}(\widehat{\zeta}_A|\zeta_A)$ is Gaussian to a good approximation but that $\mathcal{P}(\widehat{\zeta}_B|\zeta_B)$ is mildly skewed.  To assess the statistical significance of the fit we again use the $\chi^2$ distribution to weight the degrees of freedom and give an approximate p-value.

\subsection{Results with fixed DM density profiles and a generalized velocity anisotropy}

In this section we perform a model comparison between different two-parameter density models (in particular, NFW, Burkert and two profiles obtained by fixing the $\alpha,\beta,\gamma$ parameters in Eq. \ref{zhao}) and allowing freedom in the scale density $\rho_0$ and the scale radius $r_s$.
 
\begin{figure*}
	\centering
		\includegraphics[width=17cm]{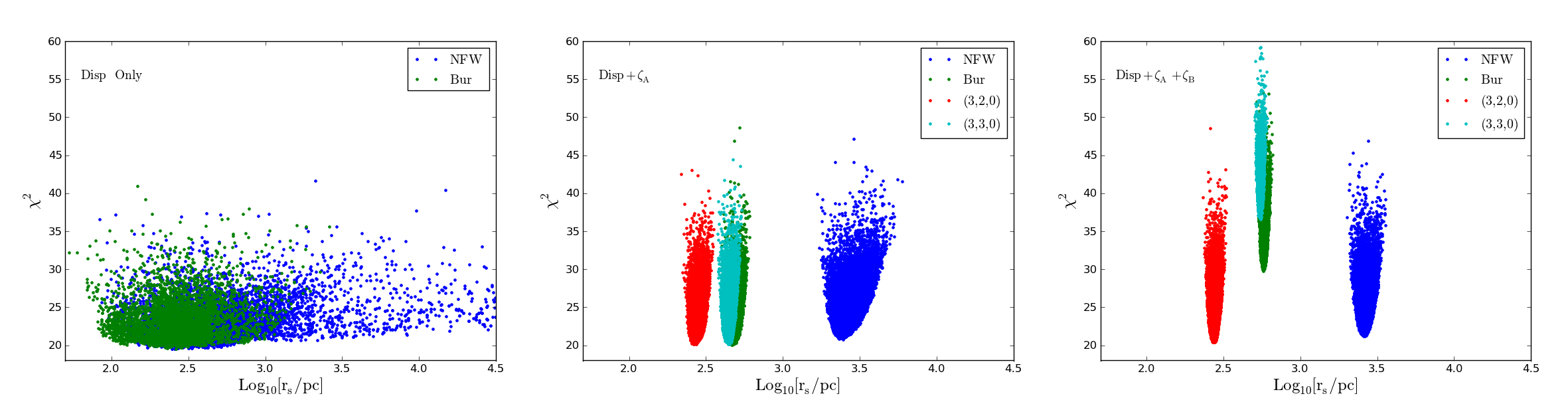}
       	\caption{$\chi^2$ values for the fits obtained as described in the text. In the panel on the left we only consider NFW and Burkert profiles and do not include fits to $\zeta_A$ or $\zeta_B$ but only the LOS velocity dispersion (26 radial bins).  In the central panel, we also fit to $\zeta_A$ while in the panel on the right we fit to velocity dispersion, $\zeta_A$ and $\zeta_B$. In the panel in the centre and on the right we also consider density profiles coming from equation \ref{zhao} with $(\alpha,\beta,\gamma)=(3,3,0)$ i.e. a highly cored profile with an NFW-like outer slope and $(3,2,0)$, a cored profile embedded within an isothermal sphere.}
\label{chiplotmain}
\end{figure*}

\subsubsection{Results with fixed Plummer profile}

In this section we again fix the tracer density of stars $\nu(r)$ to the simple one-parameter Plummer profile that was used in \S3.  Our justification for this choice is that by fitting different surface brightness profiles $\Sigma(R)$, \cite{batdat} (and see the thesis of Giuseppina Battaglia) found that for a single component of stars, Plummer models are a better fit to star counts than King, exponential and Sersic density profiles (though multiple component models provide considerably better fits than any single component model). Adopting the widely used Plummer profile also provides a useful comparison with other studies in the literature.

The main result of this work is set out in Fig. \ref{chiplotmain}.  In the left panel of this figure, we see the LOS velocity dispersion fits for NFW and Burkert profiles.  It is clear to see that with 26 radial bins, one can obtain excellent fits with both NFW and Burkert, a perfect example of the typical $\beta$ degeneracy problem in Jeans analyses. On inspection of the blue NFW points we note that despite the increased density of points below $\rm{Log}_{10}r_s = 3.2$kpc, the goodness of fit is not significantly worse at larger radii. This may suggest that the posterior distribution for $r_s$ is driven by priors rather than the likelihood function and should be treated with caution. As discussed above, we adopt the p-value approach of \cite{strigkurt}.
 
Things instantly become much more interesting when we include fits to $\zeta_A$ which we have done in the central panel, fitting it in combination with the LOS dispersion for NFW, Burkert and $(\alpha,\beta,\gamma)$ profiles corresponding to $(\alpha,\beta,\gamma)=(3,3,0)$ i.e. a cored profile with an NFW-like outer density slope and $(3,2,0)$, a cored profile with an outer density that falls like $r^{-2}$, so a cored isothermal sphere.  

For each different parametrization of the density, $\zeta_A$ focuses sharply on one particular value of the scale radius $r_s$ as can be seen in the central panel of Fig.\ref{chiplotmain}.  We can see from Fig. \ref{sculptconstep} that only a small range of $r_s$ values give curves that pass through the red square. The strength of $\widehat{\zeta}$ is that it uses the entire data set of over 1000 stars which makes the green regions very narrow.

This is particularly striking and interesting but it doesn't turn out on its own to be particularly useful in determining which functional form is the best fit as all four density profiles give rise to a comparable best $\chi^2$ (since we have fixed the values of $\alpha,\beta$ and $\gamma$ by hand for now, all four density parametrizations have the same number of degrees of freedom).  In fact, what we see is the degeneracy between the shape of the density profile and the scale radius that is discussed in \S\ref{zinterp} where the cuspy NFW model has a significantly larger scale radius than the cored models. We found that this result was robust to the choice of tracer density profile $\nu(r)$.

However when we also include the fit to $\zeta_B$ we find we are able to identify differences in the best values of $\chi^2$ for the four profile shapes.  In complete contrast to the results of \S\ref{fixbetsec} we see in the far right-hand panel that the NFW profile is now a better fit than the Burkert profile, while the alternative (3,3,0) profile is an even worse fit.  The cored isothermal sphere is actually a similarly good fit to NFW but experience from $N$-body simulations suggests an outer profile dropping as slowly as $r^{-2}$ would be difficult to understand in the context of current ideas about structure formation. Clearly, in the case where the tracer density is fixed then $\zeta_B$ (with its extra $R^2$ position-weighting) helps to distinguish between the DM models with different outer density slopes.    

\begin{figure}
	\centering
		\includegraphics[width=9cm]{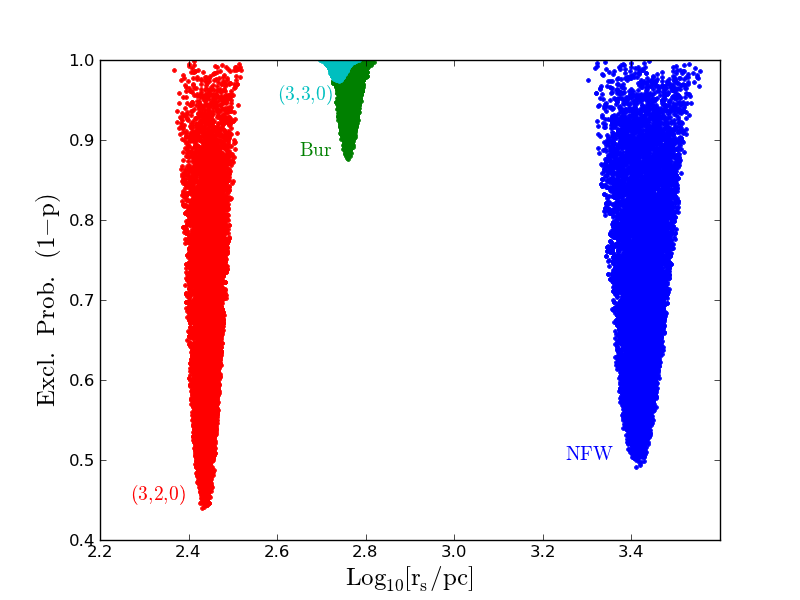}
       	\caption{To show the statistical significance of the model comparison the $\chi^2$ points in the right panel of Fig. \ref{chiplotmain} were converted to p-values. We show the exclusion probability $1-p$ which gives the fraction of the $\chi^2$ distribution which is lower than the observed value of $\chi^2$.}
	\label{probplot}
\end{figure}

To see how statistically significant the results are we converted the $\chi^2$ values to p-values with a $\chi^2$ distribution. The number of degrees of freedom is the number of dispersion bins + 2 (for both $\widehat{\zeta}$ measurements) - the number of free parameters (3 anisotropy + 2 density) -1 = 22. In Fig. \ref{probplot} we see that if the total number density of stars follows a Plummer profile, then the Burkert model can be excluded with a probability of almost 0.9 and the (3,3,0) model can be excluded with a probability of more than 0.95.

\begin{figure}
	\centering
		\includegraphics[width=9cm]{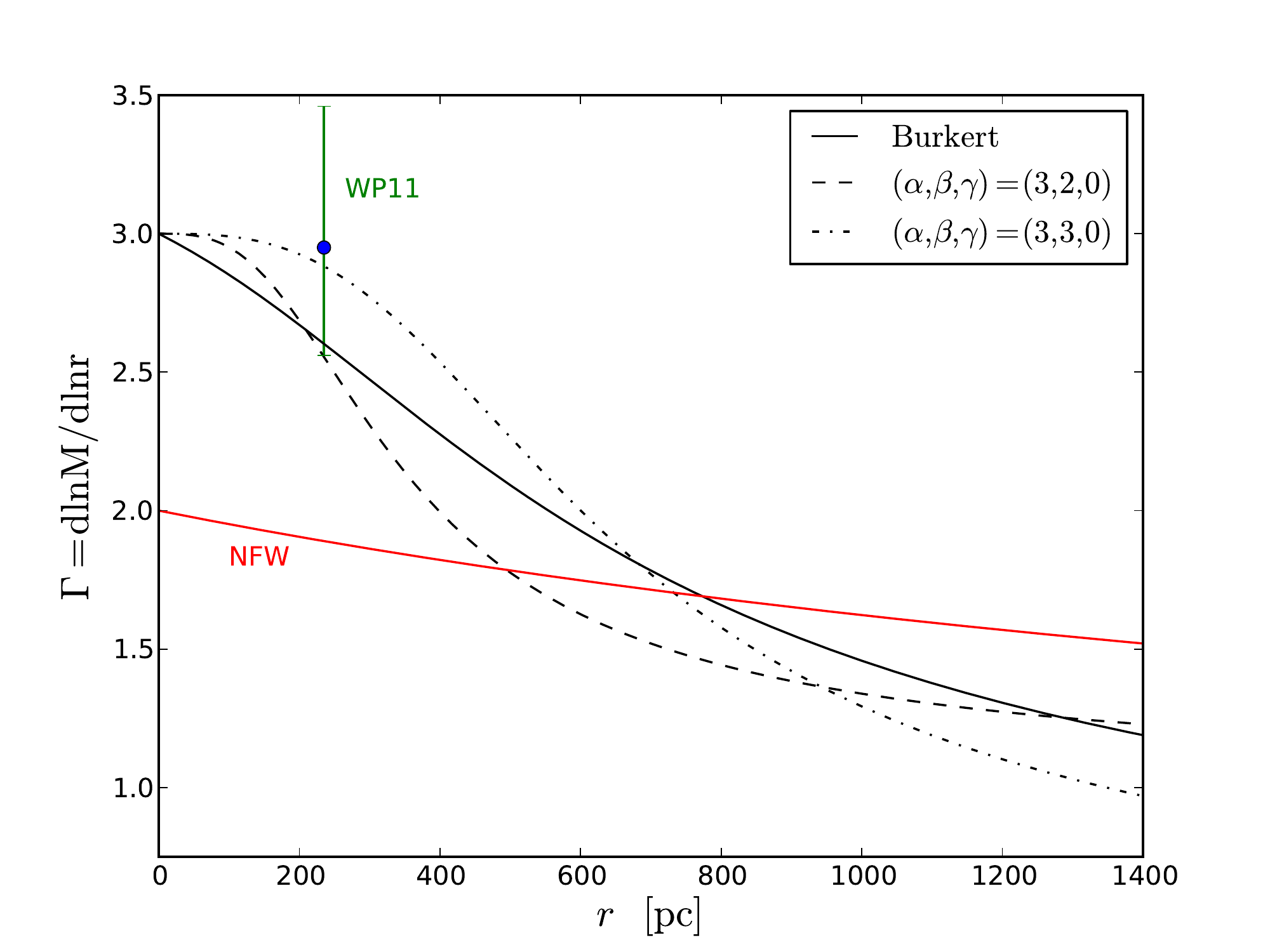}
       	\caption{For each 2-parameter density model we plot the logarithmic mass slope $\Gamma$ for the best-fitting $r_s$ parameter in the right panel of Fig. \ref{chiplotmain}. The data point labelled WP11 shows a measurement for $\Gamma$ from \protect\cite{penarrubia} which was derived using multiple stellar sub-populations.}
	\label{denprofs}
\end{figure}

For a comparison with results from the studies of multiple stellar sub-populations we plot in Fig. \ref{denprofs} the mass slope $\Gamma$ measured in \cite{penarrubia} against the best-fitting parameters for each density model. As discussed in that paper, the data point excludes NFW models to a high significance. Due to the relatively small scale radius of the cored (3,2,0) model we see that the isothermal outer density slope pulls the mass slope below the WP11 data point but remains consistent at a $1\sigma$ level. The cored (3,3,0) model with an NFW-like outer density slope is in excellent agreement with the WP11 study but as discussed previously it is ruled out in the analysis.

\begin{figure}
	\centering
		\includegraphics[width=9cm]{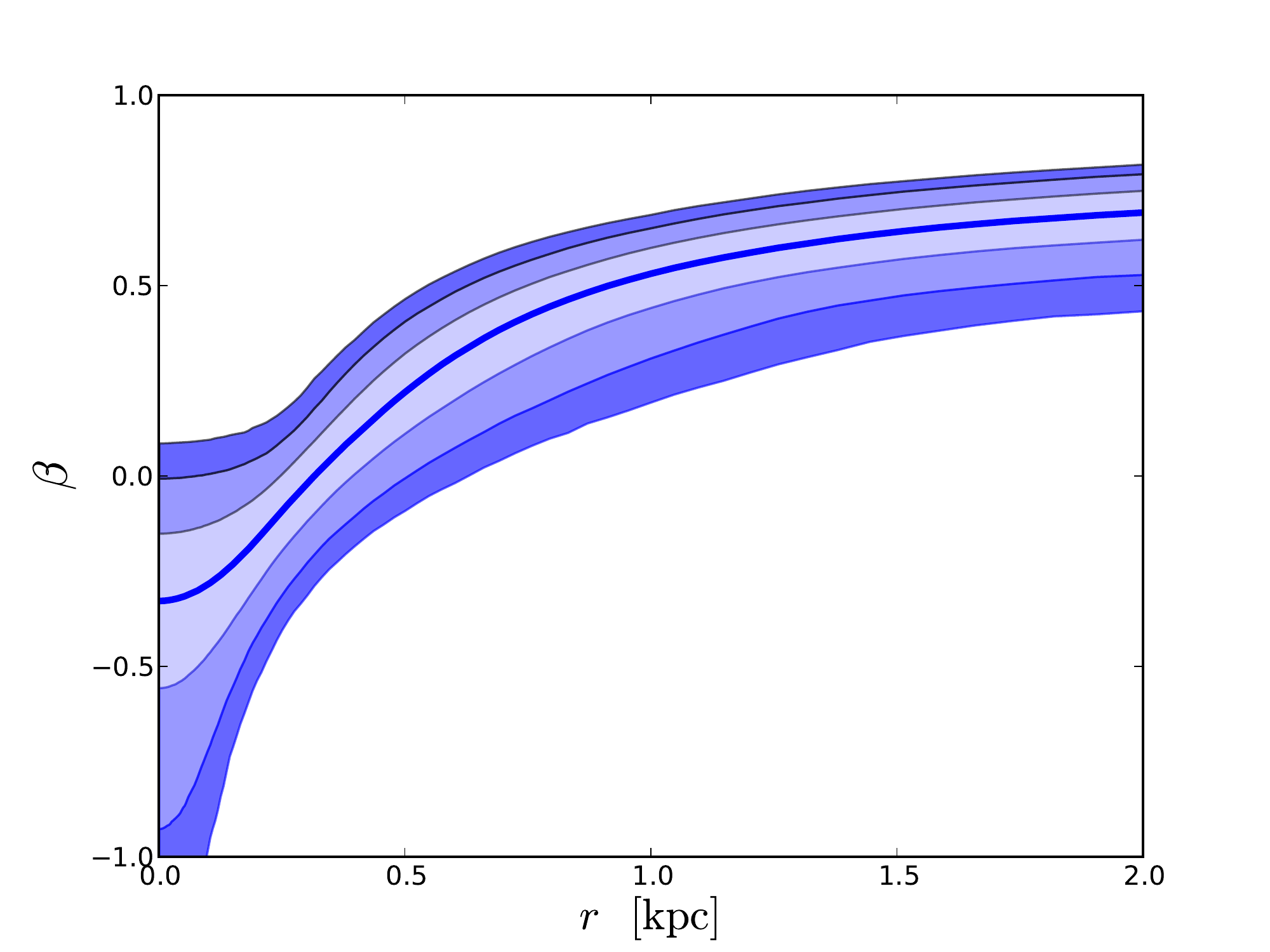}
       	\caption{The anisotropy profile in Eq. \ref{bvh} is plotted for fits to the dispersion and both $\zeta$ parameters with an NFW profile. Regions correspond to fits with p-values smaller than 0.13, 0.05 and 0.01 from lightest to darkest and the central line indicates the best fit.}
	\label{NFWbeta}
\end{figure}

 In \S\ref{fixbetsec} we saw that when $\beta=$constant NFW profiles are apparently at odds with $\zeta_A$ and $\zeta_B$.  We subsequently saw that this is completely overturned when we increase the flexibility in $\beta$.  To see why this might be the case we plot the anisotropy profile for the best fits to the dispersion and $\zeta$ parameters for the NFW with non-constant $\beta$ in Fig. \ref{NFWbeta}. All models with constant anisotropy are excluded by the $\zeta$ parameters and we see that anisotropy profiles that rise towards positive values at large radii are consistent with the LOS dispersion data. With more radial orbits at large radii the NFW models can now match the peaked velocity distributions with large values of $\zeta_A$. Interestingly, rising anisotropy profiles are sometimes seen (e.g \cite{wojbet}) for the DM particles in CDM simulations. Rising anisotropy profiles are however in stark contrast with results from an interesting orbit-based study of Sculptor \citep{breddels} which finds that the best NFW fits have tangential anisotropy. 

To understand where this discrepancy might occur we refer the reader to fig. 7 of \cite{breddels} which displays the kurtosis profile of the best fitting NFW model with tangential anisotropy. First we note that the kurtosis measured in bins 4-6 is above 3 in luminous regions so that one can reasonably expect $\widehat{\zeta}_A>3$. As discussed previously, NFW models with negative $\beta$ are not able to describe such a high value of $\zeta_A$ and we see that the best fitting solution in the figure misses each of these bins. Instead, the best fit to the kurtosis is driven by the final radial bin which receives the least weight from $\zeta_A$. The cause of discrepancy can therefore be attributed to the relative weighting of $\zeta_A$ to the most luminous regions of the galaxy which forces the fit to match the peaked distributions at around $300$pc with more radial anisotropy.

\begin{figure}
	\centering
		\includegraphics[width=9cm]{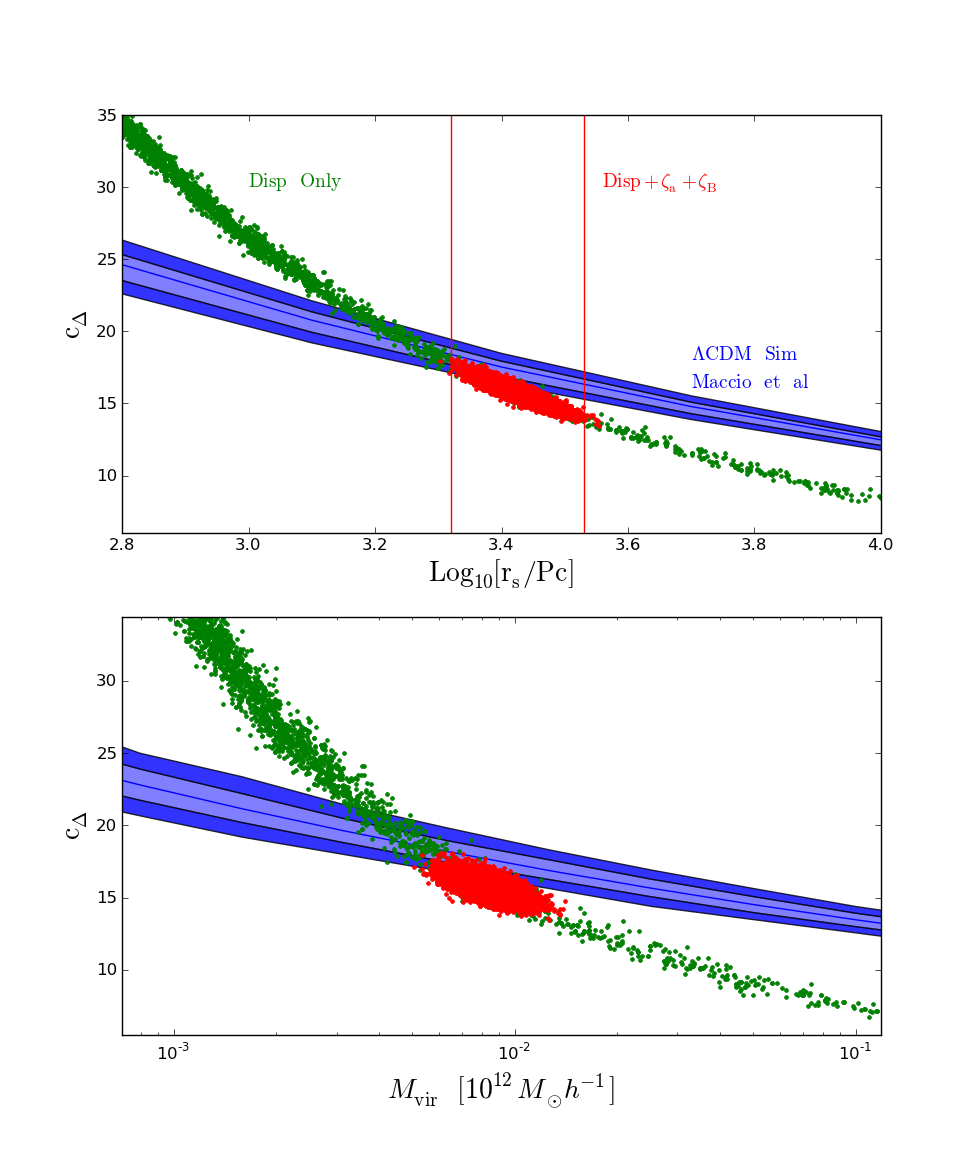}
       	\caption{In green we see the concentration parameters for the NFW profiles favoured by the LOS velocity dispersion while in red we see those that are also favoured by $\zeta_A$ and $\zeta_B$. For both fits we assume a fixed Plummer profile for the number density of stars. The $\zeta$ parameters pick out values that are in quite good agreement with $\Lambda$CDM $N$-body simulations. In the upper panel we show the concentration of the DM halo plotted against its scale radius and the blue band shows the adapted mass concentration relation in Eq \ref{cmadapt}. In the bottom panel we show the same data but plotted against the mass at the Virial radius $r_{\rm{vir}}=c_\Delta r_s$ and the blue band corresponds to the mass concentration relation in Eq \ref{cm}}.
	\label{conc}
\end{figure}

The restricted range of scale radii for NFW models also allows us to test predictions for the concentration $c_\Delta$ of DM haloes in CDM simulations. \cite{maccio} provide a mass-concentration relation that includes dwarf spheroidal sized haloes,
\begin{equation}\label{cm}
\rm{Log}_{10} c_\Delta = 1.02^{\pm}[0.015] - 0.109^{\pm}[0.005] (\rm{Log}_{10}M_{\rm{vir}}-12)
\end{equation}
where $M_{\rm{vir}}$ is the mass at the Virial radius in units of $M_\odot h^{-1}$. Noting that $M_{\rm{vir}} \propto c^3_\Delta r^3_{s}$ we can rearrange this equation to get,
\begin{equation}\label{cmadapt}
\rm{Log}_{10} c_\Delta = 2.08 - 0.246 \rm{Log}_{10}[r_{s}/\rm{pc}].
\end{equation}
For completeness, the concentration can then be related to the scale density $\rho_0$ via,
\begin{equation}
\rho_0 = \rho_{\rm{crit}} \frac{\Delta}{3}\frac{c^3}{\ln(1+c)-\frac{c}{1+c}}.
\end{equation}
 where $\Delta=98$ is the density contrast that is used in \cite{maccio} which we omit from the subscript in $c_\Delta$ for brevity. While such a test can only really be qualitative due to the uncertainties associated with computational simulations, it is interesting to note that the small range of NFW profiles picked out by the $\zeta$ parameters are in good agreement with predictions for the concentration parameters for this size of halo (see Fig. \ref{conc}). The range of Virial masses $M_{\rm{vir}}$ picked out by the $\zeta$ parameters is smaller than the spread observed in \cite{breddels}. We believe that this discrepancy is due to our additional assumption of a fixed Plummer profile for the stars. As we will see in the following section, allowing freedom in the outer slope of the stellar density profile can relax the constraining power of $\zeta_B$. Nevertheless, it is interesting that the simple Plummer+NFW model that has often been employed in the literature remains consistent with expectations from CDM simulations.

\subsubsection{Increased self-gravity and alternative tracer density profiles}

Here we discuss the sensitivity of the results presented above to the mass-luminosity ratio of the stars and the choice of the stellar density profile $\nu(r)$ that has hitherto been fixed to a cored Plummer profile.

We tested the effect of increasing the constant mass-luminosity ratio of the stars from our benchmark model of $\Upsilon_\star = 1 M_{\sun} / L_{\sun}$ to $\Upsilon_\star = 3.5 M_{\sun} / L_{\sun}$. This caused a slight increase in the best fitting scale radius for each model but made no impact on which models provided the best fit to the data. Again, it is interesting that to note the similarity with \cite{evansvirial} who find that increased self-gravity of the stars requires larger DM scale radii to fit multiple populations in Sculptor with the projected Virial theorem.  

We then experimented with different shapes for the tracer density profile within the framework of the Zhao parametrization by allowing the parameters in Eq. \ref{tracezhao} to vary freely.

 Adjusting the inner slope had little bearing on the final results which perhaps reflects the fact that both $\zeta$ parameters place little weight on stars at the centre of the galaxy. We found that though increasing the inner slope of the density profile gave rise to increased anisotropy at the centre (i.e $\beta_0$ is larger) which increases $\zeta_A$, the luminosity weighting also shifts towards the centre which causes $\zeta_A$ to fall and cancels this effect to some extent. Changing the outer density slope had little impact on $\zeta_A$ but are marked impact on $\zeta_B$ which is more sensitive to data at large radii. Increasing the outer density slope tended to increase $\zeta_B$. Crucially, boosting $\zeta_B$ relative to $\zeta_A$ is needed to increase the quality of fit for Burkert and cored DM density profiles with outer slopes steeper than an isothermal sphere. 

\begin{figure}
\centering
\includegraphics[width=9cm]{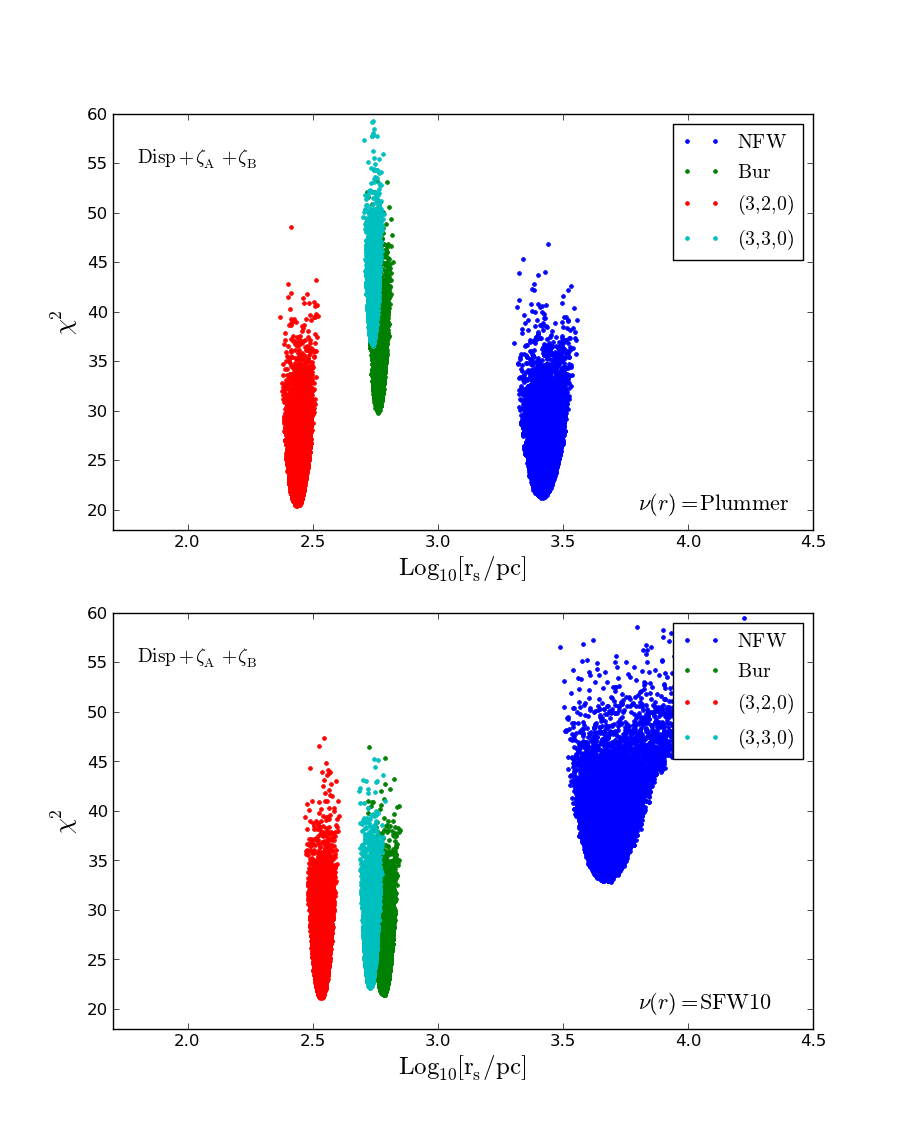}
       \caption{We repeated the analysis presented in the right hand panel of Fig. \ref{chiplotmain} with an SFW10 tracer density profile (see the text for details) that has steeper values of the inner and outer slopes than the Plummer profile which is used in the rest of this work.}   
\label{altrace}
\end{figure}

\cite{strigkurt} found that a Zhao density profile for the stellar population $\nu(r)$ with steeper inner and outer density slopes (specifically with $(\alpha_\star,\beta_\star,\gamma_\star)=(3,5.5,0.5)$ replacing the Plummer values $(\alpha_\star,\beta_\star,\gamma_\star)=(2,5,0)$) can also provide an acceptable $\chi^2$ fit to the photometry in Sculptor provided in \cite{battaglia}. We repeated the analysis presented in Fig. \ref{chiplotmain} with this new model for $\nu(r)$ which we call SFW10. In Fig. \ref{altrace} we see that even a small increase of $\Delta \beta_\star=0.5$ has a profound impact on the constraints from $\zeta_B$. As discussed above, the Burkert and (3,3,0) models can now produce larger values of $\zeta_B / \zeta_A$ to match the observation of $\widehat{\zeta}_B / \widehat{\zeta}_A$ that is observed without tuning the anisotropy parameter to values that yield poor fits to the dispersion data. We also see that scale radii favoured by $\zeta_A$ for NFW fits are now too small to fit $\zeta_B$. As such, NFW fits with the SFW10 stellar density profile have a significantly worse $\chi^2$ than the cored models.

\begin{figure}
\centering
\includegraphics[width=9cm]{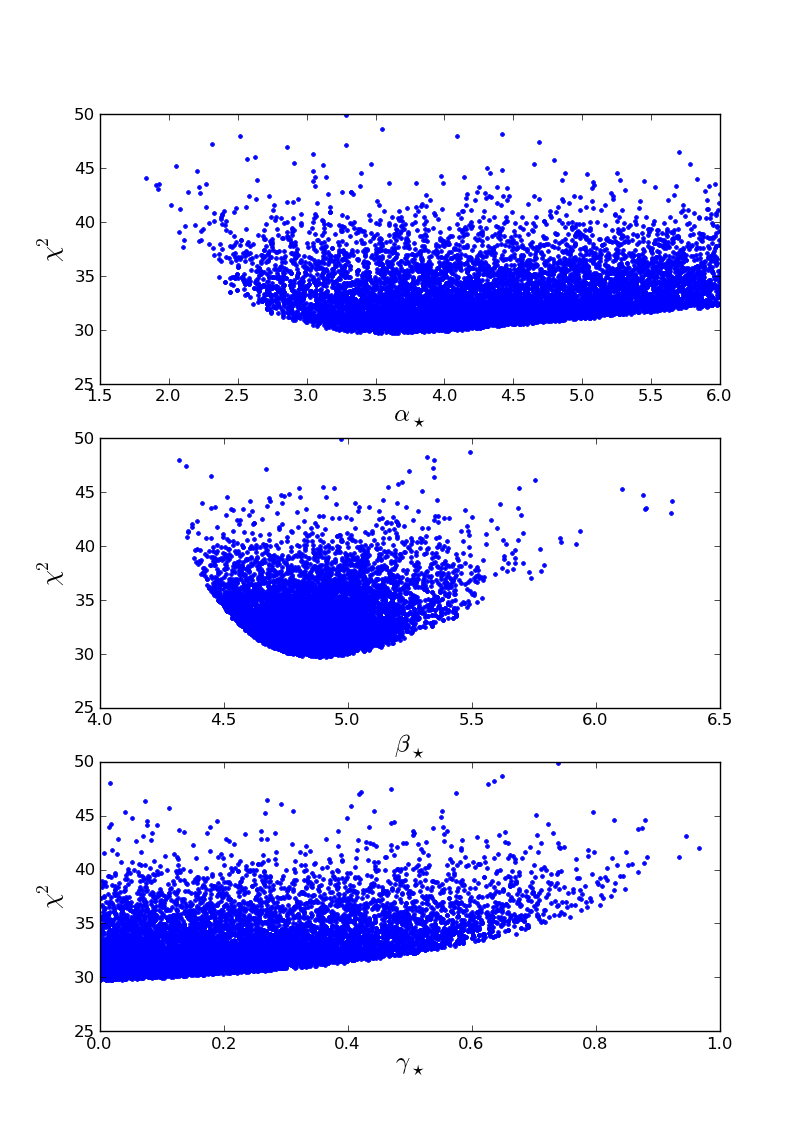}
       \caption{We fit the number of stars at each projected radius to predictions from models with a generalized Zhao profile (Eq. \ref{tracezhao}) for the number density of stars $\nu(r)$. For each model we used the Abel inversion to calculate the surface density $\Sigma(R)$ from $\nu(r)$ numerically. Displayed are points returned by MULTINEST for the shape parameters of $\nu(r)$. Here $\chi^2$ is a pseudo-chi squared goodness-of-fit statistic given by Eq. \ref{pseudochi} and the 'pseudo' refers to the fact that for a small number count the Poisson distributions used in the likelihood function do not resemble a Gaussian distribution. We split the stars into 27 radial bins spanning $50$pc so with four fitting parameters there are 23 degrees of freedom.}   
\label{traceshapepms}
\end{figure}

It is interesting to note from fig. 13 in \cite{Amorisco} that the data set used in this work appears to have a more concentrated population of tracers than data from \cite{battaglia}. For this reason we decided to perform our own fit to the star count of probable members in our sample to see how much variation we found in the outer density slope $\beta_\star$. Due to the spatial sampling bias of selecting red giant candidates for spectroscopic measurements (see \S2.3 in \cite{penarrubia}) these fits should only serve as a rough guide for the true shape of $\nu(r)$. In Fig. \ref{traceshapepms} we see that whilst Plummer-like outer slopes ($\beta_\star \sim 5$) and inner slopes ($\gamma_\star=0$) provide the best fits to the star counts, steeper outer slopes with $\beta_\star =5.5$ remain acceptable. For this reason we believe that if the parameters in $\nu(r)$ are free to vary and the star count fit is added to the likelihood function then the analysis would not be able to discriminate between Burkert and NFW models as it could in the previous section when $\nu(r)$ was fixed. The steepness of transition between the outer and inner slopes $\alpha_\star$ that we observe between the inner and outer density slopes in Fig. \ref{traceshapepms} is larger than the Plummer profile ($\alpha_\star = 2$) which could reflect the fact that two stellar sub-populations are being modelled as one. We found that altering $\alpha_\star$ made a much less significant impact on the analysis than $\beta_\star$. 

 In summary, we find that a general description of $\nu(r)$ as well as of $\beta(r)$ and (effectively) $\beta^{\prime}(r)$ is crucial in a fourth order Jeans or Virial analysis. Having approached the complete generality of spherical orbit-based modelling with our analytic approach we find a similar result to \cite{Breddelsall}; the fourth moments are not able to completely break the mass-anisotropy degeneracy but leave a space of solutions with cored density profiles degenerate with less concentrated cusped ones. Most of the power in constraining the scale radius of the DM halo comes from $\zeta_A$ which is much more robust to the choice of the tracer density profile than $\zeta_B$.


\subsection{Results with Plummer profile, generalized density profiles and generalized velocity anisotropy}
\begin{figure}
	\centering
		\includegraphics[width=9cm]{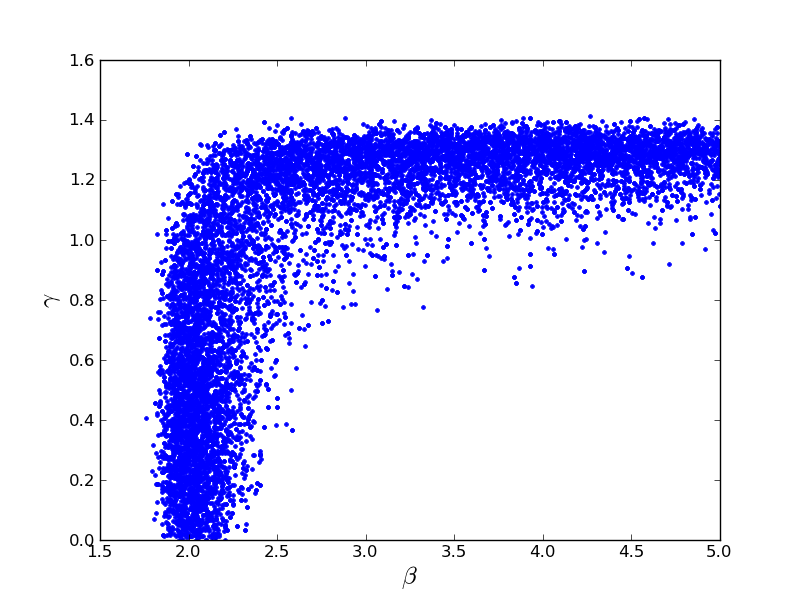}
       	\caption{Results of fitting the LOS velocity dispersion, $\zeta_A$ and $\zeta_B$ with a generalized velocity anisotropy and generalized $(\alpha,\beta,\gamma)$ density profile. Here $\beta$ refers to the outer density slope and should not be confused with the velocity anisotropy parameter. The points plotted are all within the 90\% compatible region (i.e a p-value of greater than 0.1).  It can be seen that the fits favour either a cored isothermal sphere ($\gamma=0,\;\beta \sim 2$) or cuspy profiles ($\gamma\sim1$) with a range of steeper outer slopes ($\gamma\sim 1.2$,\;$\beta>2.5$) including the NFW model with $\beta=3$.}
	\label{betgamplot}
\end{figure}
Finally we perform fits which exploit the full freedom of the Zhao $(\alpha,\beta,\gamma)$ profiles to understand the results of the last section and to see if the new results we obtain are consistent with what we discovered in the previous section. Whilst the NFW profile corresponds to a Zhao profile with $(\alpha,\beta,\gamma)=(1,3,1)$ we stress again that the Burkert profile is not nested within the Zhao parametrization. We vary $\alpha$, $\beta$ and $\gamma$ and see which values give us the best combined fit to the LOS velocity dispersion, $\zeta_A$ and $\zeta_B$. As in the previous section we assume that the stars follow a Plummer density profile. 

We saw in the previous section that both NFW profiles and cored isothermal spheres, i.e.  $(\alpha,\;\beta,\gamma)=(3,2,0)$ profiles, gave the best fit to the data with the lowest $\chi^2$ for the cored isothermal spheres slightly beating the lowest $\chi^2$ for the NFW profiles, although not by much.  At the same time, cored profiles with steeper outer density slopes like Burkert and (3,3,0) profiles appeared to be disfavoured.  In Fig. \ref{betgamplot} we see the results of the analysis.

The fits seem to favour either a cored isothermal sphere ($\gamma=0,\;\beta\sim 2$) or a (really quite) cuspy profile with a relatively steep inner and outer slope ($\gamma\sim 1.2$, $\beta>2.5$). Cored profiles with $\gamma\sim 0$ and NFW-like outer slopes seem to be very much disfavoured, a result which seems to be at odds with the results other analyses in the literature (though a recent study of Sculptor with Schwarzschild modelling \citep{Breddelsall} indicates that there is slightly (though not significantly) less Bayesian evidence for cored models with $\beta=3$ than for NFW fits and less still for cored models with $\beta=4$). Again we stress that these results depend strongly on the assumed form of the tracer density profile but it is interesting to note that a large range of outer density slopes are compatible with cusped models whilst only $r^{-2}$ slopes are accepted for cored models.

\section{Spherical symmetry and dynamic equilibrium}

The two fundamental assumptions that underpin the analysis in this work (and indeed any spherical Jeans/Schwarzschild method based on the Collisionless Boltzmann Equation) are that (i) the distribution function of the tracers $f(r,\textbf{v})$ and gravitational potential $\Phi(r)$ are spherically symmetric and (ii) the tracer population has reached equilibrium and $\partial_t f = 0$.

$N$-body simulations of dwarf galaxies orbiting the MW indicate that the average effect of tidal stripping (over all projected angles) is to inflate the LOS dispersion at large radii \citep{readtides}. This can lead \citep{klimenttidesouter} to an observer overestimating the halo mass (and in particular the scale radius) and/or underestimating the anisotropy parameter $\beta$.

The final radial bins in our sample (with 26 bins) show that though the LOS dispersion rises in the interior of the galaxy it begins to decline at $R=1$kpc which suggests that tidal stripping may not be significant below this radius.

Be that as it may, if the tidal tails are oriented along the LOS to dwarf spheroidal galaxies then it has been argued \citep{klimenttidesorient} that tidal effects can dramatically alter the kinematics. Accurate proper motion measurements for Sculptor from the \textit{Gaia} satellite will provide more information about the orbit of Sculptor and the feasibility of this scenario. As noted in \cite{evansvirial}, current astrometric measurements of Sculptor's proper motion with the Hubble Space Telescope (HST) \citep{piatek} indicate that Sculptor has not passed the pericentre of its orbit for $~1$Gyr and that a pericentric distance of $69$kpc makes it unlikely that tidal effects will ever be strong enough to make a significant impact.      

Evidence for mild asphericity in Sculptor is irrefutable with an observed ellipticity of $e\sim 0.3$ in the stellar surface profile. Though one might expect that the dark matter halo is more spherical than the embedded tracer population one must also factor in the unknown LOS elongation. This arguably presents the greatest challenge to modelling the dynamics of dwarf spheroidal galaxies. The problem can be approached in two ways (i) develop increasingly sophisticated modelling techniques to fit highly accurate simulations of triaxial haloes and (ii) use simple analytic methods to gain a deeper intuitive understanding of dynamics in spherical systems and look for observational signatures that are robust to the deviations from sphericity or guide the modelling of aspherical systems.

Some success for the latter approach comes from the surprising finding that the mass-slope method of \cite{penarrubia} may be robust when applied to triaxial haloes \citep{chervin}. With such simple analytic methods, the authors found that to some extent the bias from asphericity effectively cancels. These results are disputed in \cite{kowal} however, who find that there is a significant bias on mass slope estimates that depends on the unknown angle between the longest axis of the stellar component and the LOS. 

\cite{evansvirial} were also able to show that their results obtained with the projected Virial theorem are insensitive to an axisymmetric flattening of the stellar population and dark matter halo.

Unfortunately, the method presented here also depends on the assumption that the velocity ellipsoid is spherical. This introduces an additional bias to the anisotropy parameter as discussed in \cite{Wojtak} for tracers in cluster sized haloes. Replacing the Jeans equations with the Virial equations enables us to include information on the shape of the LOS velocity distribution without reference to $\beta^{\prime}$, the anisotropy between the shape of the radial and tangential distributions. This is a positive step towards modelling higher moments of the LOS velocity distributions in a way that is robust to the spherical assumption.

\section{Conclusions}

In this work we have discussed the use of higher order Virial equations rather than higher order Jeans equations to try and break the degeneracies associated with the traditional Jeans analysis. The main advantage of this approach is that the fourth order Virial equations add constraints to the fourth moment of the velocity distribution without adding new fitting parameters to the standard Jeans analysis that can compromise efforts to break the $\beta$ degeneracy.  We used the fourth order Virial equations to derive two new parameters $\zeta_A$ and $\zeta_B$ which replace the kurtosis as a measure of the shape of the LOS velocity distribution. Each solution to the second order Jeans equation could then be checked for consistency with shape information from the data with simple estimators $\widehat{\zeta}$. In contrast to the kurtosis, these global estimators sum over all stars in the data set which removes the issue of binning and maximizes the statistical precision of the fourth order moments.

In a simplified analysis with a constant velocity anisotropy parameter, we have shown that it is clear how the $\zeta$ parameters reduce the space of solutions  when they are added to a standard Jeans equation fit.  We found that velocity anisotropy primarily affects the ratio $\zeta_A/\zeta_B$ and that  both $\zeta$ parameters are highly sensitive to the scale radius of the DM halo.  For fixed anisotropy and DM scale radius, cored DM profiles have higher values of both $\zeta$ parameters than cusped haloes which suggests that they have more peaked velocity distributions on average. 

When we generalise the velocity anisotropy parameter to allow it to be a function of radius, we see that again the $\zeta$ parameters focus in on particular solutions, most strikingly, for a given parametrization, $\zeta_A$ focuses in on a particular value of the scale radius but does not favour any particular density profile shape relative to any other (cored or cusped).  There is a degeneracy between cored profiles and cusped profiles with larger scale radii. 

Including $\zeta_B$ in the analyses 
can break this degeneracy to some extent and places strong constraints on the outer slope of the dark matter density profile in particular. In the case where $\nu(r)$ was fixed to a Plummer profile we found that the $\widehat{\zeta}$ measurements for Sculptor could rule out cored profiles that are common in the literature such as the Burkert profile that fits low surface brightness spiral galaxies \citep{spiralcores,salucciwalker} or cored profiles with NFW-like outer slopes $(\alpha,\beta,\gamma)=(3,3,0)$ and suggest a smaller mass slope than is predicted with multiple stellar populations in \cite{penarrubia}. However, whilst $\zeta_A$ is relatively robust to the choice of number density for the stars $\nu$, the extra $R^2$ position-weighting in $\zeta_B$ makes it highly sensitive to the outer slope. When we increased the outer slope of $\nu(r)$ to $\beta_\star = 5.5$ we found that both Burkert and (3,3,0) models could be made consistent with $\widehat{\zeta}$. A fit to the star counts with a general form for $\nu(r)$ showed that whilst Plummer models are preferred, there is enough freedom to allow a steeper outer slope.

In short, we find that when one considers (i) a single tracer component, (ii) all possible fits to the tracer density $\nu(r)$ and (iii) complete freedom for the anisotropy parameter $\beta(r)$ to vary within the limits of its parametrization and without assuming anything whatsoever about the fourth order shape anisotropy parameter $\beta^{\prime}(r)$, then the fourth moments alone are not often able to solve the cusp/core degeneracy. The fourth order Virial equations isolate information from the fourth order Jeans equation that is independent of $\beta^{\prime}$ and we therefore believe that a fourth order Jeans analysis that satisfies the criteria listed above would not provide significantly more information \footnote{An interesting correlation between the two anisotropy parameters observed in \cite{wojbet} would break the $\beta^{\prime}$ degeneracy however.}. This was also the conclusion presented in \cite{Breddelsall} with orbit-based modelling and interestingly they also find that the remaining set of degenerate solutions has less concentrated cusped haloes and more concentrated cored ones. We believe that our simple analytic method brings this result into sharp focus.

Our results contrast with conclusions presented in previous work that found density profiles with extended cores to be favoured in a fourth order Jeans analysis of Sculptor. To understand this discrepancy we first note that the Einasto density profile chosen for the dark matter in that work does not allow an independent description of the inner and outer density slopes. With $\nu(r)$ fixed to a Plummer profile it is conceivable that the tight constraints placed on the outer slope of the density profile by the fourth order Jeans equations (which implicitly include $\zeta_B$) were responsible for the discriminating power of the analysis. This highlights the importance of considering highly generalized or non-parametric profiles for the tracer density and anisotropy parameters. The Virial equations presented in this work are valid for any physical choice of $\nu(r)$ and $\beta(r)$ and remove the need to parameterize $\beta^{\prime}(r)$. 

Though the fourth moments do not specify a single density profile, they can be used to make very tight predictions for the velocity anisotropy and concentration of density profiles with fixed shape parameters. In fact, given that the $\zeta$ parameters place such tight constraints on NFW models it is remarkable that both the concentration and anisotropy parameter is in reasonable agreement with expectations from $\Lambda$CDM simulations. The tight constraints on the concentration are powered primarily by $\zeta_A$ alone which is relatively robust to the choice of the tracer density profile. Interestingly, the results presented in \cite{evansvirial} and \cite{amorfornax} suggest that whilst cored density profiles provide the best fit, NFW profiles with sufficiently large scale radii (and in the case of Sculptor lower mass-luminosity ratios for the stars than $\Upsilon_\star = 8$) may also fit the energetics of multiple stellar sub-populations in Fornax and Sculptor implied by the projected Virial theorem. The authors argue however, that the necessary scale radii for NFW fits imply DM haloes that are not concentrated enough to match predictions from $\Lambda$CDM simulations. In contrast to those studies and in line with \cite{breddels}, we find sufficient scatter in the mass-concentration relation to accommodate NFW models in our analysis. It would be interesting to see whether this holds for other dwarf spheroidal galaxies and the $\zeta$ parameters can provide a stern test for the concept of a universal density profile (see e.g \cite{penuni} and \cite{jardelnonpar} for other recent challenges). 

Studies with multiple populations continue to provide the strongest constraints on the inner slope of DM density profiles. With the new and simpler method introduced in this paper it is much more feasible to introduce fourth moments for each individual population. With two separate strong predictions for the DM scale radius from two separate $\zeta_A$ measurements, the fourth moments could provide a powerful complement to the usual multiple population mass estimators.

A better understanding of the density profiles in dwarf spheroidal galaxies, or indeed any galaxies, is extremely valuable as it can help to understand the effect of baryons upon dark matter profiles.  For any given density model, the dramatic improvement afforded by the $\zeta$ parameters in recovering the scale radius could be extremely useful for calculating the magnitude of the astrophysical J-factor and therefore obtaining constraints on the self annihilation of dark matter. The fact that the method gives much tighter constraints on the underlying radial functions of $\beta$ may potentially yield information about structure formation history.

Two important challenges to this work that could become a problem are the validity of assuming spherical symmetry and dynamic equilibrium for Sculptor.  We feel that by reducing the higher moment analysis to its simplest and most robust form to date we have made a positive step towards testing these assumptions.    

Finally, we emphasize that our discussion of the new Virial shape parameters has no bearing on the scale of the system and can be readily applied to any spherical  system of tracers in a gravitational potential.  Whenever the standard Jeans equation is applied, the Virial shape parameters offer two new extra constraints on the system for free without introducing any new parameters.
 

\section*{Acknowledgements}
In order to perform this work, we have made extensive use of simulated data from the \textit{Gaia} challenge data sets provided by Matthew Walker and Jorge Pe\~{n}arrubia which were critical to our analysis.  The authors are also grateful to Justin Read and the University of Surrey for hosting the first \textit{Gaia} Challenge where very useful progress was made.  We are also thankful for the comments of an anonymous referee on a previous paper.  TR thanks the KCL Graduate School for support.  MF is grateful to the STFC for funding.

\bibliographystyle{mn2e2}
\bibliography{tranrep}
\appendix

\section{Systematics on Dispersion and $\widehat{\zeta}$ data}

In this section we discuss how numerous potential sources of systematic error can affect the observed values of the LOS dispersion and $\zeta$ parameters. Where possible we use the \textit{Gaia} Challenge test suite to estimate these systematics numerically and simulations including effects such as binary motion, experimental velocity errors, perspective rotation and probability of membership are all generously provided by Matthew Walker and Jorge Pe\~{n}arrubia. As the true $\zeta$ parameters and LOS dispersion are known for each model in the test suite we can estimate the total bias $\zeta - \widehat{\zeta}$ and fit to it as shown in Figs. \ref{gaiaepA} and \ref{gaiaepB}. As discussed in the text, experimental errors can be understood by performing a convolution of the intrinsic velocity distribution with a Gaussian error distribution. The more complicated contribution of other systematics to the bias is outlined in greater detail below.

\subsection{Binary stars}

Binary fractions for Sculptor have been estimated in the literature \citep{Queloz} at around $f_b \sim 0.2-0.5\%$. Very recently, \citep{minor} performed a study on the variability of repeat velocity measurements in the Sculptor data set \citep{walkdat} that we consider in this work. In the study, a likelihood analysis was performed to assess to what extent the variability of repeat velocity measurements can be attributed to experimental velocity errors or to binarity. The findings suggest that Sculptor possesses a relatively large binary fraction of $f_b = 0.59$ that is similar to the binary fraction that is found in metal poor Milky Way field star populations \citep{minor}. This fraction is based upon the assumption that the mean log period of binary stars in dwarf spheroidals is similar to that in MW field stars. A smaller mean log period would suggest a smaller binary fraction. Without a detailed knowledge of how e.g star formation affects binary fractions in dSphs relative to the field, we cannot be sure that this (or indeed the converse scenario) is the case. Nevertheless, in accordance with \cite{minor} we adopt a binary fraction of $f_b=0.6$ in our analysis.

\begin{figure}
\centering
\includegraphics[width=9cm]{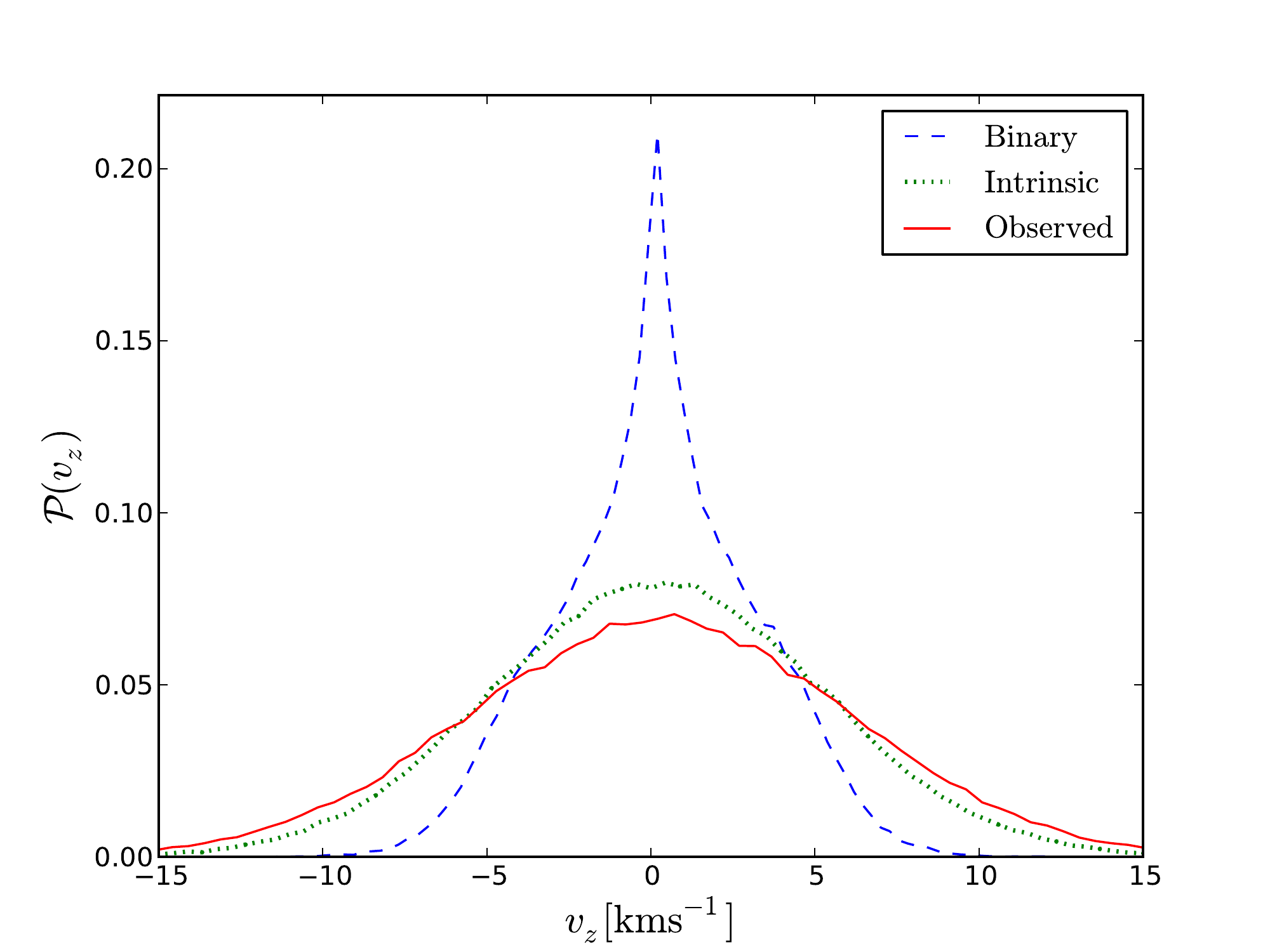}
       \caption{LOS velocity distributions for the binary motion (blue dashed), intrinsic motion (green dotted) and observed motion (red solid). Here we assume that the intrinsic velocity distribution is Gaussian with a dispersion of $5 \rm{kms}^{-1}$. This is then convolved with the binary LOS velocity distribution (see text for details) to give the LOS velocity distribution that would be observed if $100\%$ of stars are binary stars.}
\label{binaries}
\end{figure}

Numerous studies (e.g \cite{binaryinflate}) have shown that a significant fraction of binary stars can artificially inflate the observed LOS velocity dispersion. Though the relatively large intrinsic LOS velocity dispersion in Sculptor reduces the impact of binary stars, the $\zeta$ variables depend on the fourth moment of the LOS velocity distribution and it has been argued \citep{dejonghebinary} that the LOS kurtosis is particularly sensitive to binary motion. To study the effects of binary motion on the $\widehat{\zeta}$ estimates we adopt LOS velocities from the \textit{Gaia} challenge test suite that follow the prescription in \cite{mcconachie}. The resulting velocity distribution for binary motion along the LOS is shown as a blue dashed curve in Fig. \ref{binaries}. We note that there is additional weight in the shoulders of the distribution relative to the distribution presented in fig. 4 of \cite{dejonghebinary}. The origin of this discrepancy requires a more detailed study but is not too surprising given the very different methodologies presented in the two papers. Different choices for parameters of the binary system (such as the mass ratio of the primary and secondary stars) could also contribute to any differences.

Despite the odd shape of the binary velocity distribution (blue dashed curve in Fig. \ref{binaries}) we found that the sample kurtosis is close to the Gaussian value of 3. Remarkably, we found that a convolution of the Gaussian intrinsic distribution with the binary distribution had a negligible impact on the shape of the observed velocity distribution but just gave a small boost to the LOS dispersion. In Fig. \ref{binaries} we see that even an intrinsic LOS velocity distribution (green dotted curve) with a small dispersion (approximately half that in Sculptor) has an almost identical shape to the observed velocity distribution (red solid curve). Moreover, we found that contrary to the results presented in \cite{dejonghebinary}, intrinsic velocity distributions more peaked than Gaussian were flattened when binary velocities were added. For these peaked models ($\zeta_A > 3$) the binary stars act like the Gaussian experimental velocity errors to create the negative bias ($\widehat{\zeta}_A - \zeta_A<0$) that we see on the right-hand side of Fig. \ref{gaiaepA}.
     
\subsection{Milky Way interlopers}

Another key issue when using higher moments of the velocity distribution is that Milky Way contaminants with large LOS velocities will impact strongly on the tails of the velocity distribution and inflate the measured kurtosis. For this reason most interloper removal schemes conservatively remove all LOS velocities that are more than $3 \sigma$ from the mean velocity of the dwarf spheroidal galaxy. Whilst this is a robust way to ensure that the LOS velocity dispersion is not artificially inflated by outliers, the removal of member stars in the tails of the distribution can severely bias the measured kurtosis. As we have shown in \S\ref{zetasec}, velocity distributions that are more peaked than Gaussian are generic predictions for models with cored dark matter density profiles, deeply embedded tracer populations or radial ($\beta>0$) anisotropy. We therefore chose to use probabilities from the method described in \cite{walkermetal} that uses velocity, position and (crucially) metallicity, to assess the likelihood of membership for each star, rather than the arguably more robust $3 \sigma$ cutoff.

 One benefit of using the luminosity weighted average is that the kurtosis-like variable $\zeta_A$ naturally weights the stars in the inner regions that are most likely to be members of the dwarf spheroidal. Unfortunately $\zeta_B$ is sensitive to cutoffs in both the LOS velocity and the projected radius $R$ which together form the dominant source of bias that we observe in Fig. \ref{gaiaepB}. For models in the \textit{Gaia} challenge test suite with low values of $\zeta_B$ we see a positive bias ($\widehat{\zeta}_B - \zeta_B >0$) whilst for $\zeta_B>5$ we see that the bias becomes negative. This can be understood by considering that models with low $\zeta_B$ naively have flatter velocity distributions and are thus principally affected by the cutoff in the projected radius $R$ which from Eq. \ref{zetaestB} increases $\widehat{\zeta}_B$. As $\zeta_B$ increases, cutting off the longer tails of the velocity distribution becomes the dominant effect which decreases $\widehat{\zeta}_B$ by reducing $v^4_z$ faster than $v^2_z$. This source of negative bias is also present for $\zeta_A$ and adds to the negative bias from binary motions and experimental errors that we see in Fig. \ref{gaiaepA}. 

We tested to see how much imposing a $3 \sigma$ cutoff (after removing stars with membership probability less than 0.95) would affect the results. After iteratively removing stars with velocities greater than $3 \sigma$ we found that as expected the kurtosis-like estimator fell from $\widehat{\zeta}_A = 3.43$ to $\widehat{\zeta}_A = 3.21$ (with a corresponding change in $\zeta_B$ from 3.69 to 3.83). The fact that $\widehat{\zeta}_A$ remains significantly (given the small errors in $\widehat{\zeta}_A$) above the Gaussian value of 3 after removing the tails of the distribution suggests that it is the shoulders of the distribution and the more pronounced central peak that are driving the positive non-Gaussianity. This has been confirmed in a study \citep{Amorisco} of Sculptor with Gauss-Hermite moments that are less susceptible to the tails of the velocity distribution. For both the data set used in this work \citep{walkdat} and an independent survey \citep{battaglia} the authors find that the Gauss Hermite moment $h_4$ is positive at all radii and maximally so in the most luminous regions. We therefore believe that the $3 \sigma$ cutoff is not appropriate for use on the Sculptor data set and that, qualitatively, the results presented here are robust to the choice of interloper removal scheme.   

\subsection{Rotation}

To assess the impact of rotation on our results we calculated the estimator $\widehat{\zeta}_A$ before and after removing the mild gradients of -5.5 km s$^{-1}$deg$^{-1}$ in the direction $\theta_{\rm{PA}} = 21 \deg$ as prescribed in \cite{walkerrot}. We found no significant difference with $\Delta \widehat{\zeta}_A \sim 0.03$. Again, because rotation primarily affects stars at large projected radii, the luminosity weighted average in $\zeta_A$ makes it less sensitive to intrinsic rotation than the Jeans equations. More surprisingly, we found that  $\zeta_B$ was not significantly affected either.

There is still debate in the literature about the systemic proper motion of Sculptor and various astrometric measurements give conflicting results. If the systemic proper motion observed with the HST \citep{piatek} is accurate then the observed velocity gradients cannot be attributed to the dwarfs systemic motion \citep{walkerrot} but suggest intrinsic rotation. Evidence for intrinsic rotation is found in another study of Sculptor's kinematics \citep{battaglia}.            

We transformed the heliocentric velocities to the dwarf rest frame by assuming the alternative proper motion measurements from the HST. We found that $\widehat{\zeta}_A$ was not significantly affected but that $\widehat{\zeta}_B$ was increased from 3.69 to 3.80 which is consistent with intrinsic rotation boosting the velocity measurements in the highly weighted outer regions.

\section{Choice of radial bins and comparison with other Sculptor data sets}

In our benchmark model we split the stars equally amongst 26 radial bins to maximize the radial coverage of the dispersion profile whilst preserving a sample of size of $N>50$ stars in each bin. Other studies in the literature employ fewer radial bins so for a direct comparison we repeated our analysis with 12 radial bins. 
\begin{figure}
\centering
\includegraphics[width=9cm]{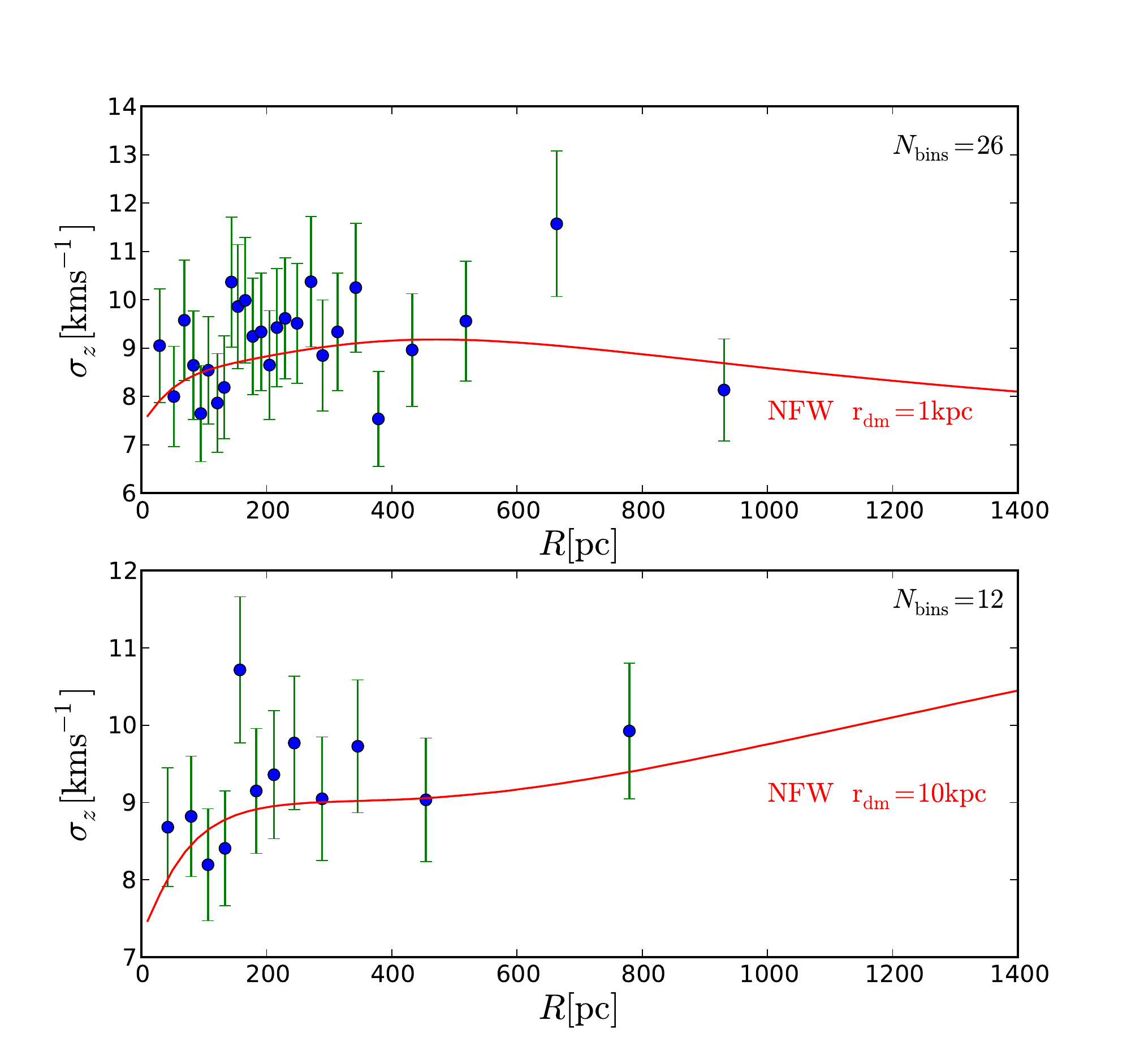}
       \caption{LOS dispersion data for binning schemes with the stars split equally into 26 (upper panel) and 12 (lower panel) radial bins. We show fits with NFW profiles to demonstrate that more extended and diffuse haloes ($r_s = 10\rm{kpc}$) have a rising dispersion profile that fits the LOS dispersion data better when the number of bins is limited to ($N_{\rm{bins}}=12$).}  
\label{binfit}
\end{figure}

In Fig. \ref{binfit} we see that a choice of 12 radial bins does not detect a slight decline that is observed in the LOS dispersion at around $R=1$kpc. Consequently, we see that a better fit can be achieved to NFW haloes with very large scale radii (and therefore lower concentrations) that produce rising LOS dispersion profiles. We also note that an alternative data set for Sculptor from the VLT/FLAMES survey \citep{battaglia} has a LOS dispersion that rises out to a greater radius $R$ than the data set that we consider in this work (see fig. 14 in \cite{Amorisco} for a direct comparison). We would therefore expect results from the alternative data set to bear a closer resemblance to the results obtained with 12 radial bins. 

\begin{figure}
\centering
\includegraphics[width=9cm]{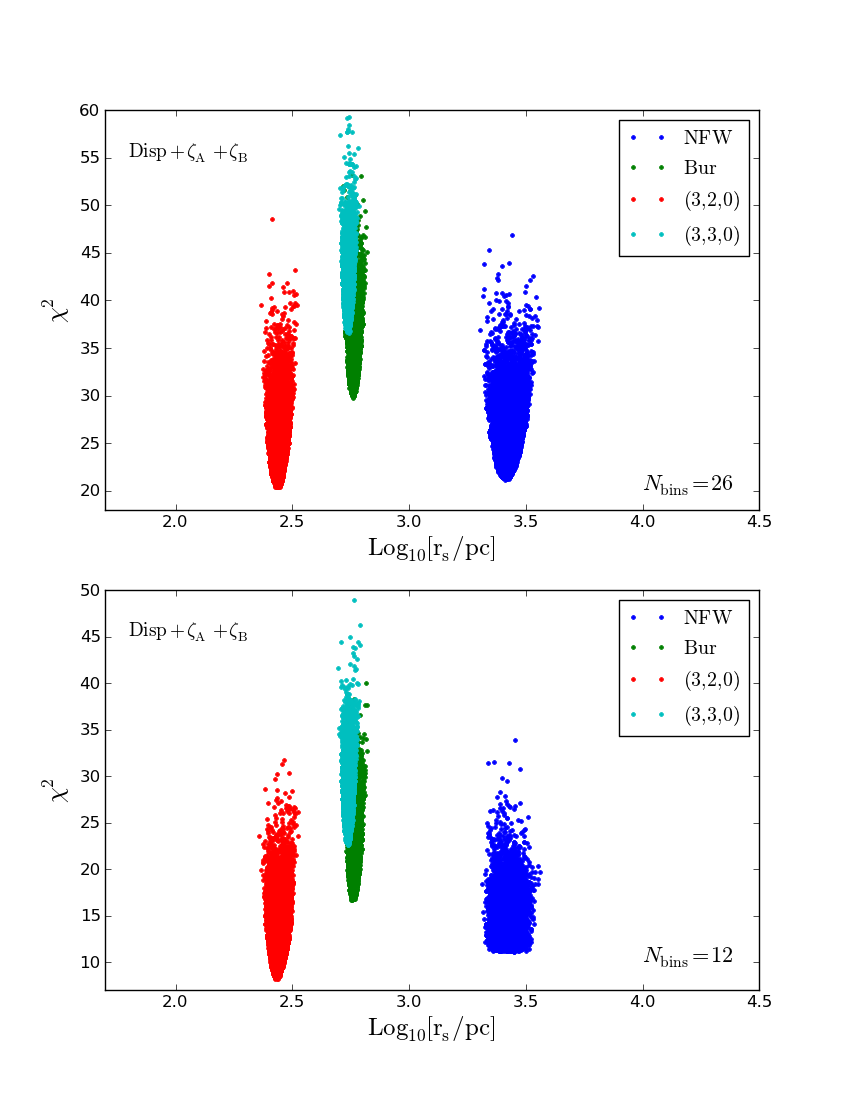}
       \caption{We repeated the analysis presented in the right hand panel of Fig. \ref{chiplotmain} with 12 radial bins for the LOS dispersion data (bottom panel) rather than 26 radial bins (upper panel) which is used in the rest of this work.}   
\label{binmodelcomp}
\end{figure}

Whilst, the choice of binning had an impact on the dispersion-only analysis (e.g. in the constant $\beta$ analysis it improves the $\chi^2$ for scale radii above $r_s = 1$kpc in Fig. \ref{constbetchi}), the $\zeta$ parameters place tight enough constraints to override these differences in the joint analysis. In Fig. \ref{binmodelcomp} we see that the relative fit between different models is largely unchanged.

\end{document}